\documentclass[useAMS,usenatbib]{mn2e}

\usepackage{Times}

\usepackage{amssymb}
\usepackage{graphicx}

\title[Two-point Correlation Function of WMAP 9 year data]{Two-point Correlation Function of WMAP 9 year data}

\author[A.Gruppuso]
{\parbox{\textwidth}{A.~Gruppuso$^{1,2}$\thanks{E-mail:gruppuso@iasfbo.inaf.it}}\vspace{0.4cm}\\
\parbox{\textwidth}{$^{1}$INAF-IASF Bologna, Via P. Gobetti 101, I-40129, Bologna, Italy\\
$^{2}$INFN, Sezione di Bologna, Via Irnerio 46, I-40126 Bologna, Italy\\
}}

\begin{document}

\date{Accepted ???. Received ???; in original form ???}

\pagerange{\pageref{firstpage}--\pageref{lastpage}} \pubyear{2012}

\maketitle

\label{firstpage}

\begin{abstract}
Using an optimal estimator for the CMB Angular Power Spectra we compute the temperature two-point correlation function of WMAP 9 year at low resolution.
Supported by realistic Monte-Carlo simulations, we evaluate how such observed function depends on the Galactic mask.
We find that it is more and more consistent with zero (i.e. no correlation) as the Galactic mask is increased. 
In particular we estimate that such a behavior happens very rarely in a $\Lambda$CDM model, i.e. $<0.01\%$ of the realizations when we adopt a mask which leaves $46 \%$ of observed sky.
This is evaluated for the so called $S_{1/2}$ estimator, already well known in literature \citep{Spergel:2003cb}. 
Also for its generalization to the whole angular range $[0, \pi]$, namely $S_{1}$,
we find a very unlikely behavior which is $ \lesssim 0.04\%$ C.L. for the considered masks that cover at least $\sim 54\%$ of the sky. 
\end{abstract}


\section{Introduction}
\label{intro}
%
Cosmic microwave background (CMB) data have greatly contributed to the building of a cosmological model, named concordance $\Lambda$ cold dark matter ($\Lambda$CDM) model, 
see \citep{Hinshaw:2012fq} for the last cosmological analysis of the Wilkinson Microwave Anisotropy Probe (WMAP) data and see \citep{Ade:2013xsa} for the recent {Planck} cosmological results.
This model involves a set of basic quantities for which CMB observations and other cosmological and astrophysical data-sets 
agree\footnote{See \citep{Ade:2013lmv} for a tension concerning $\Omega_m$ extracted from { Planck} CMB and galaxy clusters data.}: spatial curvature close to zero; 
$\sim68.5 \%$ of the cosmic density in the form of Dark Energy; $\sim26.5\%$ in Cold Dark Matter (CDM); $\sim5\%$ in baryonic matter; and a nearly scale invariant
adiabatic, Gaussian primordial perturbations \citep{Ade:2013lta}.

However there are several interesting deviation from the $\Lambda$CDM model, often called anomalies, specially at large angular scales \citep{Copi:2010na} where CMB anisotropies probe the 
physics of the early universe. If they are statistical flukes or deterministically due to some unknown effect is still an open question. 
See \citep{Copi:2013zja} for a prescription based on polarization data, to test the hypothesis that the large-angle CMB temperature perturbations in our Universe represent a rare statistical fluctuation within
$\Lambda$CDM model. See \citep{Bennett:2010jb} for a discussion about the ``a posteriori'' bias that might affect these analyses. 

In the current paper we focus on the lack of power in the two point correlation function of the temperature CMB anisotropies for angles larger than $60^{\circ}$.
Such intriguing discrepancy has been already noted with COBE data \citep{Hinshaw:1996ut} and then by the WMAP team in their first year release \citep{Spergel:2003cb}.
In \citep{Copi:2006tu, Copi:2008hw} it is shown that this event happens in only 0.03\% of realizations of the $\Lambda$CDM model using WMAP 3 and 5 years data.
Such a lack of power is confirmed in a later analysis \citep{Efstathiou:2009di} using WMAP 5 year data but at the same time it is found with a Bayesian approach that the $\Lambda$CDM model
cannot be excluded.
WMAP 7 year data are taken into account by \cite{Sarkar:2010yj}, where it is also shown that such anomaly does not correlate with the anomalous alignment of the $\ell=2$ and $\ell=3$
multipoles. 

Here, we compute the two-point correlation function, $C (\theta)_{TT}$, using WMAP 9 year low resolution data in temperature. 
This correlation function is defined as
\begin{equation}
C(\theta)_{TT}= \sum_{\ell \ge 2}^{\ell_{max}} \xi_{\ell} P_{\ell}(\theta) \, C_{\ell}^{TT} \, , \label{CTT}
\end{equation}
where $\xi_{\ell}= {(2 \ell +1) / {4 \pi}} $, $P_{\ell}$ are the Legendre polynomials and
with $C_{\ell}^{TT}$ being the angular power spectrum (APS) of the temperature CMB map.
We build $C (\theta)_{TT}$ through Eq.~(\ref{CTT}) evaluating the APS with a quadratic maximum likelihood (QML) estimator. 
This method is proven to be optimal since it provides unbiased and minimum variance estimates \citep{Tegmark:1996qt,Tegmark:2001zv,Gruppuso:2009ab}. 
The optimality of the QML method is compared to pseudo-$C_{\ell}$ methods in \citep{Efstathiou:2003tv}. 
In \citep{Molinari2013} it is shown that at the lowest multipoles (i.e. $\ell \lesssim 20$) the variance of the 
QML method is roughly half that of the pseudo-$C_{\ell}$ approach. 
This makes the QML method for APS essential for the computation of the two point correlation function since such an object is dominated by the lowest multipoles.

Once the two-point correlation function is computed, we evaluate the following estimator, $S_{1/2}$ \citep{Spergel:2003cb}
\begin{equation}
S_{1/2} = \int_{\pi/3}^{\pi} d \theta \, \left( C (\theta)_{TT} \right)^2 \sin \theta \, ,
\label{s1su2estimator}
\end{equation}
as well as its natural generalization $S_{1}$ to the whole angular range
\begin{equation}
S_{1} = \int_{0}^{\pi} d \theta \, \left( C (\theta)_{TT}  \right)^2 \sin \theta \, .
\label{sfullestimator}
\end{equation}
Eqs.~(\ref{s1su2estimator}) and (\ref{sfullestimator}) have to be considered as estimators of the distance from the null value. 
They are used to test the lack of correlation, i.e. how much likely is for a CMB extraction (compatible with the WMAP 9 best fit model) to be close to the zero value.
Instead in order to test the compatibility with the $\Lambda$CDM model we define the following analogous estimators
\begin{equation}
S_{1/2}^{\Lambda} = \int_{\pi/3}^{\pi} d \theta \, \left( C (\theta)_{TT}  - C(\theta)_{TT}^{\Lambda} \right)^2 \sin \theta \, ,
\label{s1su2estimatorLambda}
\end{equation}
\begin{equation}
S_{1}^{\Lambda} = \int_{0}^{\pi} d \theta \, \left( C (\theta)_{TT}  - C(\theta)_{TT}^{\Lambda} \right)^2 \sin \theta \, ,
\label{sfullestimatorLambda}
\end{equation}
where $ C(\theta)_{TT}^{\Lambda}$ is the two point correlation function for temperature CMB anisotropies expected in a given $\Lambda$CDM model,
that in following will be the WMAP 9 best fit model. 
Eqs.~(\ref{s1su2estimatorLambda}) and (\ref{sfullestimatorLambda}) have to be considered as estimators of the distance from $C(\theta)_{TT}^{\Lambda}$. 
They are used to test the compatibility with the WMAP 9 best fit model.

Supported by realistic Monte Carlo simulations we evaluate Eqs.~(\ref{s1su2estimator}), (\ref{sfullestimator}),  (\ref{s1su2estimatorLambda}), (\ref{sfullestimatorLambda})
and compare with WMAP 9 year data.
Moreover we test the stability of our results on various Galactic sky cuts. 

The paper is organized as follows.
Section \ref{dataset} is devoted to the description of the considered WMAP 9 year low resolution data set. 
A general analysis of the two-point correlation function of WMAP 9 year data is given in Section \ref{twopoints}.
The evaluation of the estimators and corresponding analysis are presented in Section \ref{estimators}.
In Section \ref{connectionwithAPS} the low amplitudes of the lowest APS are recognized as responsible of the lack of correlation at large scales.
This makes a connection with the Low Variance anomaly, see for example \citep{Monteserin:2007fv,Cruz:2010ud,Gruppuso:2013xba}.
Conclusions are drawn in Section \ref{conclusions}.
In Appendix \ref{QMLBolpol} details about the APS estimator are provided.
Appendix \ref{consistency} gives a comparison between the estimators build with our APS extractor and with the spectrum provided by the WMAP team  \citep{Bennett:2012zja}.


\section{Data set}
\label{dataset}
We use the temperature ILC WMAP 9 year map, available at the LAMBDA website\footnote{http://lambda.gsfc.nasa.gov/}, smoothed at $9.1285$ degrees and reconstructed at HealPix\footnote{http://healpix.jpl.nasa.gov/}
 \citep{gorski} resolution $N_{side}=16$. 
We have added to that map a random noise realization with variance of $1 \mu K^2$ as suggested in \citep{Dunkley:2008ie}.
This is done to regularize the inversion of the covariance matrix.
Because of its amplitude, such an additional white noise covers the correlated noise present in the ILC map due to the smoothing of the data
and, at the same time, is sufficiently low to not impact the subsequent analysis. 
Consistently, the noise covariance matrix for TT is taken to be diagonal with variance equal to $1 \mu K^2$ when using our QML implementation, namely {\sc BolPol} \citep{Gruppuso:2009ab}.
See also Appendix \ref{QMLBolpol}.

\begin{figure}
\centering
\includegraphics[width=80mm]{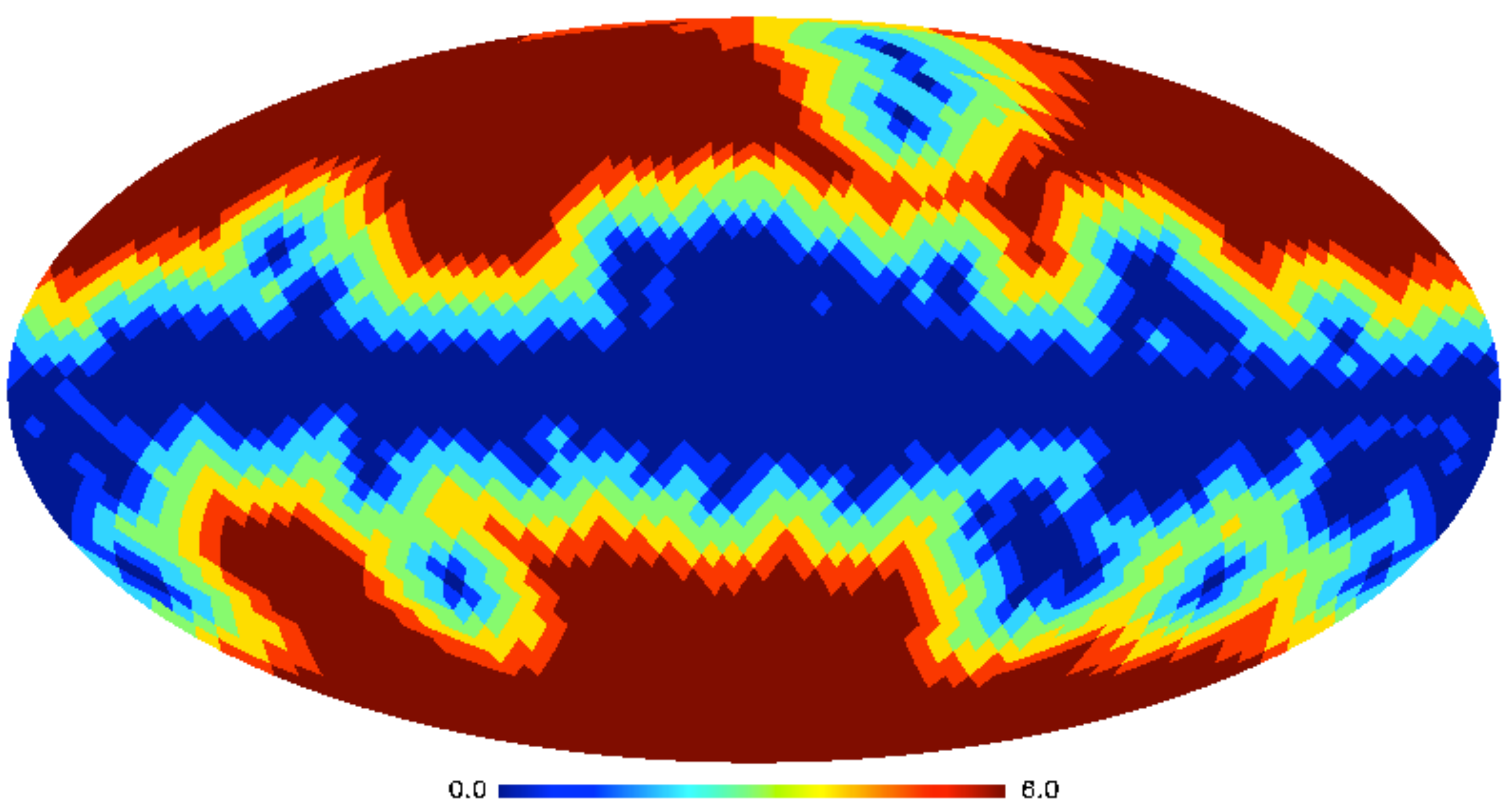}
\caption{Galactic temperature masks. 
Dark blue region is for case ``a''. Dark blue and blue regions are for case ``b''. Dark blue, blue and light blue regions are for case ``c''.
Dark blue, blue, light blue and green regions stand for case ``d''.
Dark blue, blue, light blue, green and orange regions represent case ``e''. 
Dark blue, blue, light blue, green, orange and light red regions are for case ``f''. 
See also Table \ref{maskstabel}.}
\label{one}
\end{figure}

The temperature ILC WMAP 9 year map has been masked with various Galactic masks that are shown in Fig. \ref{one}.
More specifically, these masks are built extending the edges of the kq85 temperature mask 
by 4, 8, 12, 16 and 20 degrees.
See Table \ref{maskstabel} for details of the considered cases including the observed sky fraction.

\section{Two-point correlation function}
\label{twopoints}

In this Section we evaluate Eq.~(\ref{CTT}) 
using 
{\sc BolPol} already employed in \citep{Gruppuso:2009ab,Paci:2010wp} for WMAP 5 year data analysis, in \citep{Gruppuso:2010nd,Gruppuso:2011ci,Paci:2013gs} for WMAP 7 year data
and in \citep{Gruppuso:2013xba} for WMAP 9 year data. See also \citep{Planck:2013kta} for an application of such a code to {Planck} data.
See Appendix \ref{QMLBolpol} for details about the QML method.
\begin{table}
\centering
\caption{Sky fraction observed with the considered masks. See also Fig.~\ref{one}.}
\label{maskstabel}
\begin{tabular}{ccc}
\hline
Case & Extension & Observed \\
 & wrt kq85 ($^{\circ}$) & sky fraction \\
\hline
a & +0 & 0.78  \\
b & +4  & 0.68  \\
c & +8  & 0.56   \\
d & +12  & 0.46  \\
e & +16  & 0.36  \\
f & +20  & 0.28  \\
\hline
\end{tabular}
\end{table}

Supported by realistic Monte-Carlo (MC) simulations we compute Eq.~(\ref{CTT}) replacing the TT APS up to $\ell_{max}=32$ for all the cases of Table \ref{maskstabel}.
This is performed in order to study the stability of such a function against the Galactic masking.
With ``realistic'' simulations we mean a set of CMB plus noise realizations where the signal is extracted from the WMAP 9 year best fit model and the noise through a Cholesky decomposition of the noise
covariance matrix. 
The resolution and the smoothing used in the simulations are of course the same as in the ILC WMAP 9 year map (i.e. $N_{side}=16$ and $FWHM=9.1285$).
We have then computed the APS by means of {\sc BolPol} for each of the simulations and for each of the cases given in Table \ref{maskstabel}. 
See right column of Table \ref{percentages} for the number of simulations, $N_{sims}$, considered in each case.

\begin{figure*}
\centering \begin{minipage}[l]{.35\textwidth}
\centering 
\includegraphics[width=74mm]{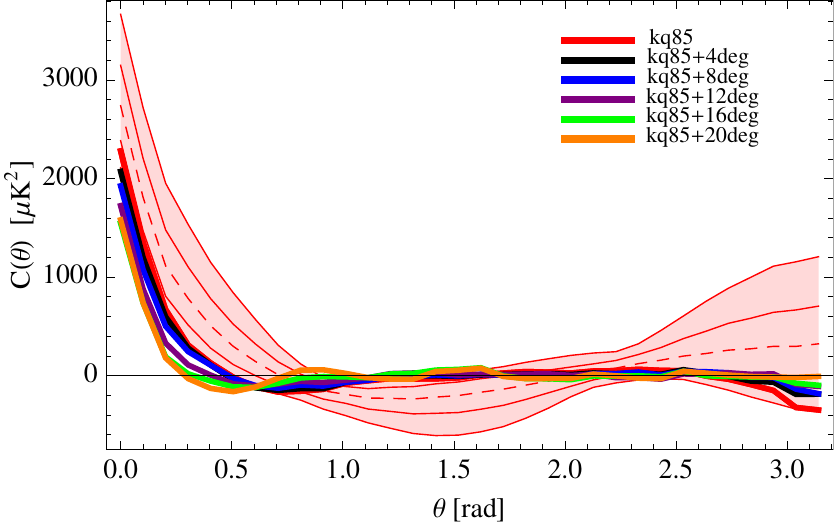} 
\end{minipage}%
\hspace{10mm}%
\begin{minipage}[r]{.55\textwidth}
\centering 
\includegraphics[width=45mm]{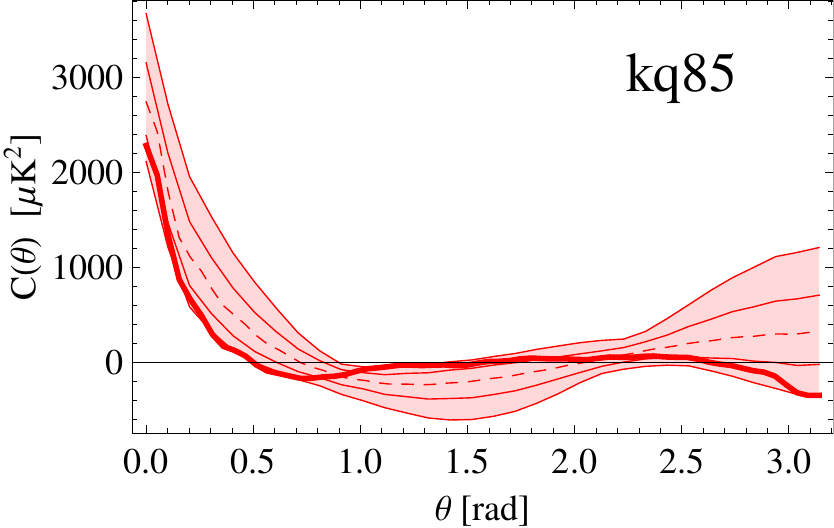}
\includegraphics[width=45mm]{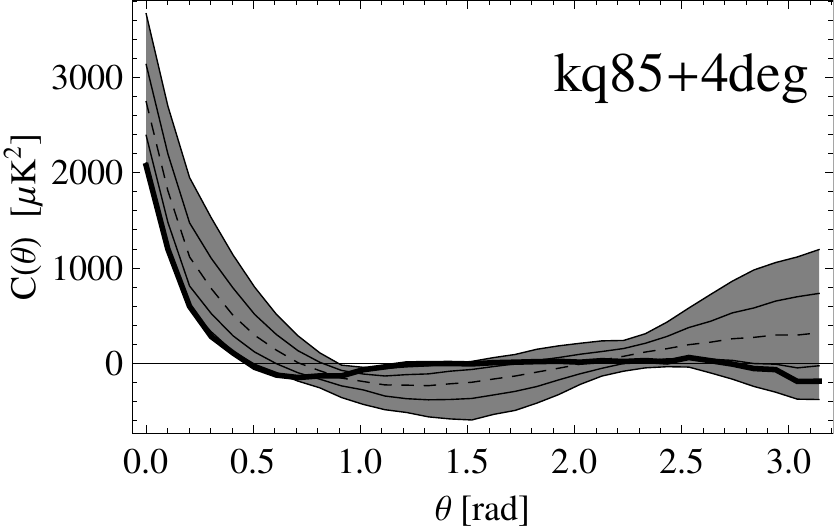}
\includegraphics[width=45mm]{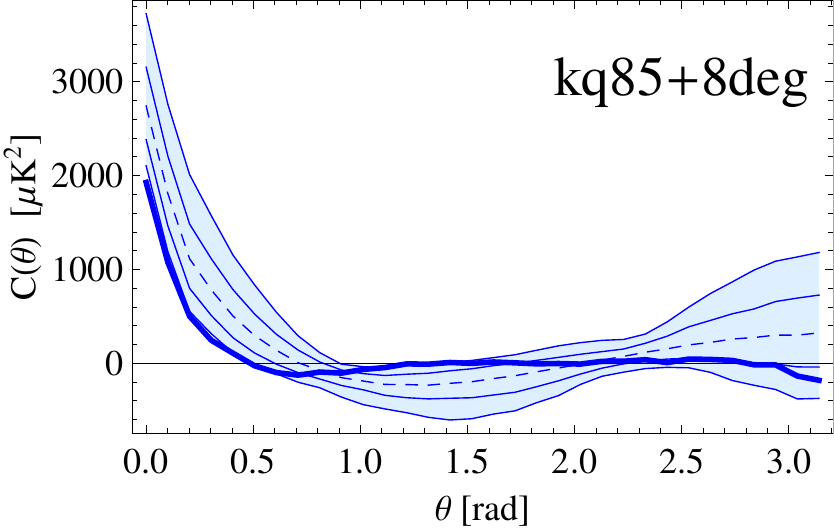}
\includegraphics[width=45mm]{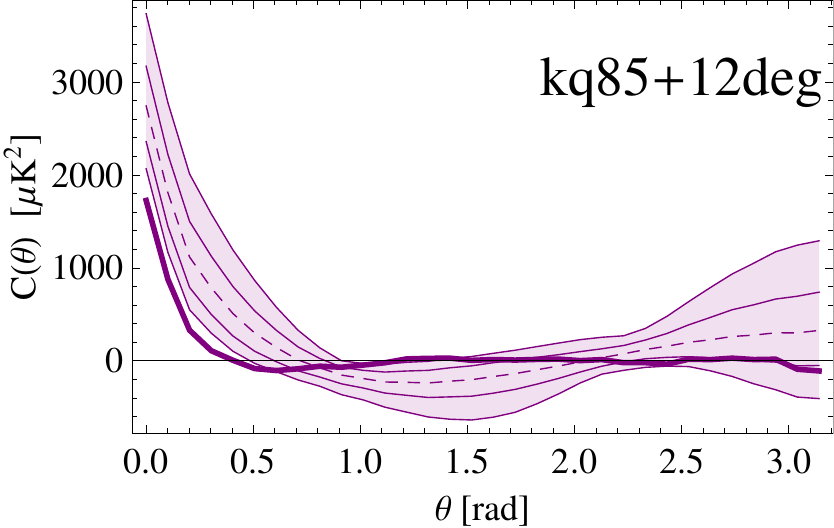}
\includegraphics[width=45mm]{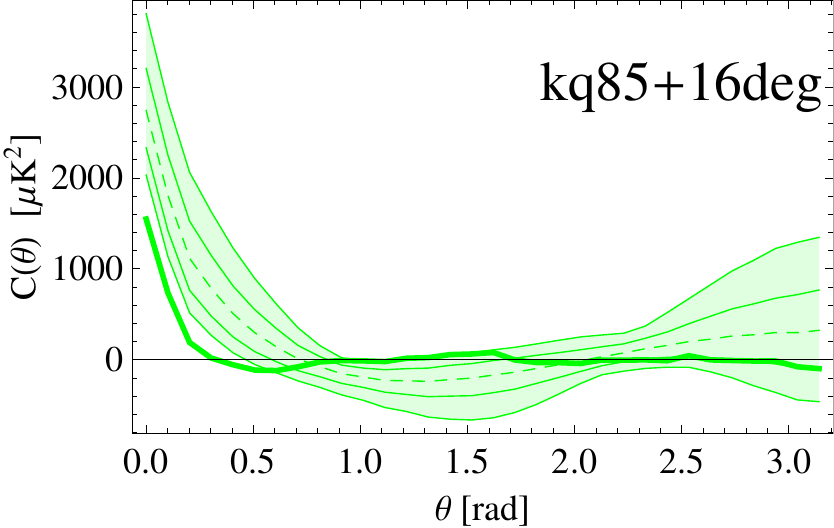}
\includegraphics[width=45mm]{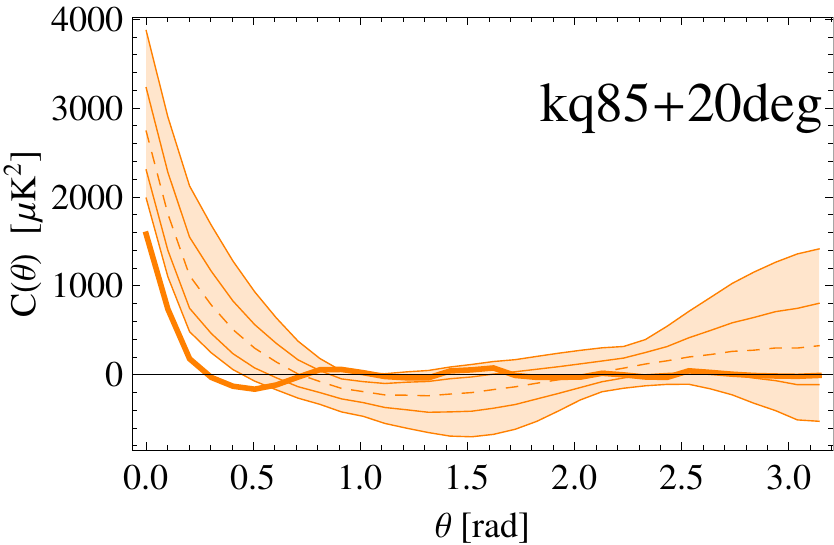}
\end{minipage} 
\caption{Left panel: TT two point correlation function for case ``a'', ``b'', ``c'', ``d'', ``e'' and ``f''. 
The red shaded area is for the 1- and 2-$\sigma$ dispersion of the $\Lambda$CDM model (dashed line) for case ``a'', and the thick solid lines are for the WMAP 9 year data.
Right panels: TT two point correlation function for case ``a'', ``b'', ``c'', ``d'',  ``e'' and ``f'' from upper left to lower right panel.
The colored shaded area is for the 1- and 2-$\sigma$ dispersion of the $\Lambda$CDM model (dashed line) for corresponding case.
Units: $\mu$K$^2$ (y-axis) and radiants (x-axis) in all the panels. 
\label{fig:minipage2}} \end{figure*}
%
%
%
Results are shown in Fig.~\ref{fig:minipage2}. 
In each panel of Fig.~\ref{fig:minipage2} the shaded area represent the 1- and 2-$\sigma$ dispersion of the $\Lambda$CDM model (dashed line) and the thick solid line is for WMAP 9 year data.
Left panel of Fig.~\ref{fig:minipage2} is for a direct comparison among the two point correlation functions
whereas right panels are for the comparisons of each case of Table \ref{maskstabel} with the corresponding MC simulations.

Three considerations stem from Fig.~\ref{fig:minipage2}. First, results of \citep{Gruppuso:2013xba} are obviously qualitatively recovered at $\theta=0$ since $P_{\ell}({\theta=0})=1$
because the two point correlation function at that value is nothing but the variance.
Second, at large angles, i.e. $\theta > 60^{\circ}$, we confirm the anomalous lack of power (see e.g. \cite{Copi:2006tu}). 
At the same time we note that at the largest scales, WMAP data approach the $\Lambda$CDM model when the mask is increased.
This will be properly quantified in Section \ref{estimators} which represents the focus of this paper.
Third, since $P_{\ell}(\theta=\pi)=(-1)^{\ell}$, we can rewrite $C(\pi)_{TT}$ as
\begin{equation}
C(\pi)_{TT} = \sum_{\ell, \,even} \left( {{2 \ell +1} \over {4 \pi}} \right) \, C_{\ell}^{TT} - \sum_{\ell, \, odd} \left( {{2 \ell +1} \over {4 \pi}} \right) \, C_{\ell}^{TT} \, .
\label{correlazioneTT} 
\end{equation}
This means that the two-point correlation function at $\theta=\pi$ is a natural estimator of the even-odd symmetry TT spectrum (often called TT Parity symmetry
\citep{Kim:2010gf,Kim:2010gd,Gruppuso:2010nd}).
Therefore Fig.~\ref{fig:minipage2} is also showing that increasing the mask, the asymmetry of the power between even and odd multipoles is 
decreasing\footnote{The TT Parity analysis is beyond the scope of this manuscript. We intend to return to this point in a separated paper.} \citep{Kim:2010st}.

See Appendix \ref{consistency} for a comparison between the two-point correlation function build with the APS estimated by {\sc BolPol} and with the spectrum provided by the WMAP team  \citep{Bennett:2012zja}.
This is done for case ``a''.

\section{Estimators}
\label{estimators}
\begin{figure*}
\includegraphics[width=43mm]{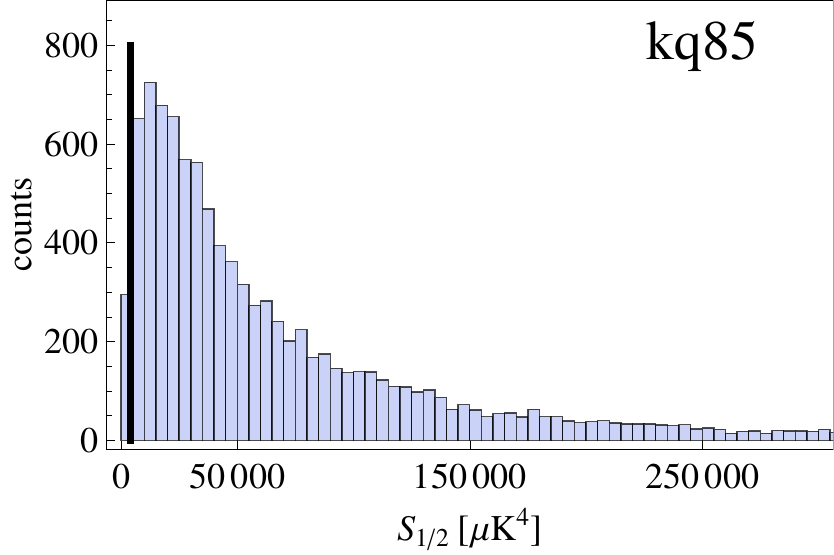}
\includegraphics[width=43mm]{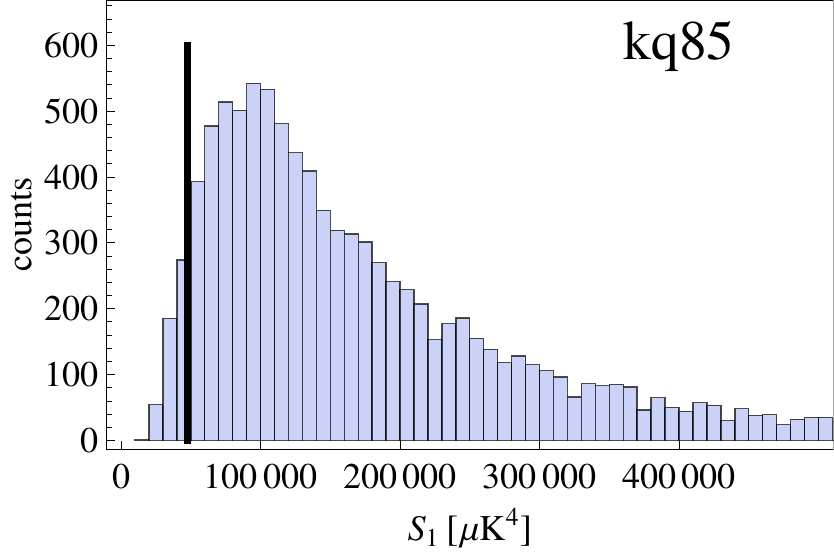}
\includegraphics[width=43mm]{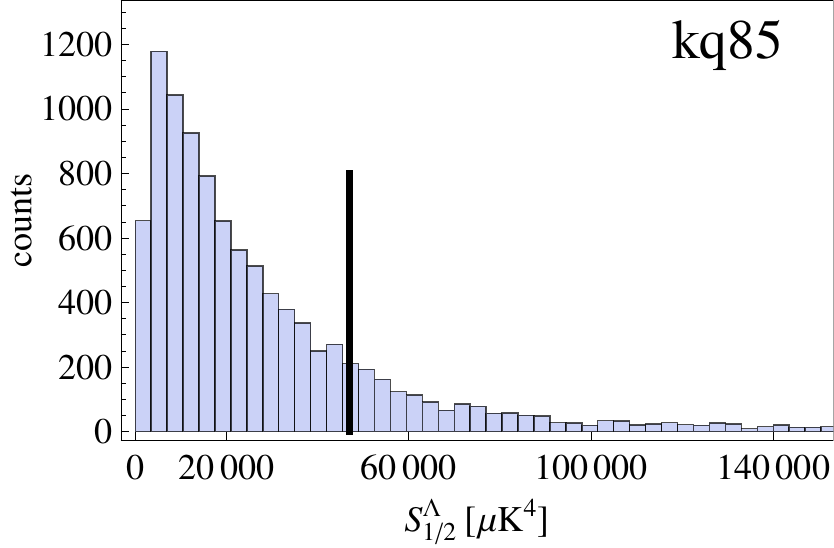}
\includegraphics[width=43mm]{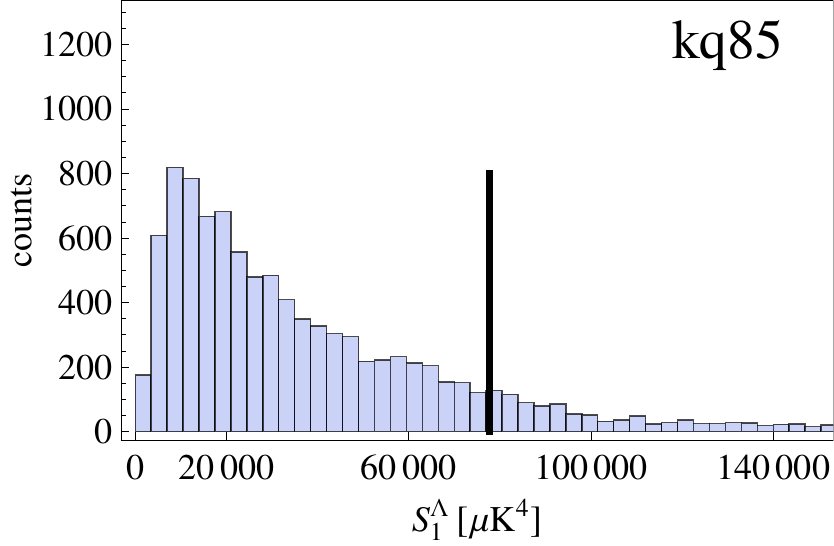}

\includegraphics[width=43mm]{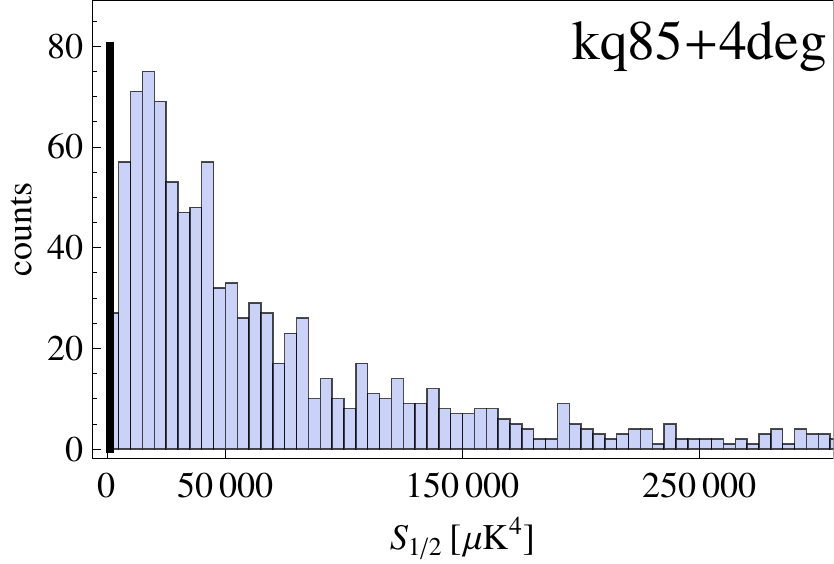}
\includegraphics[width=43mm]{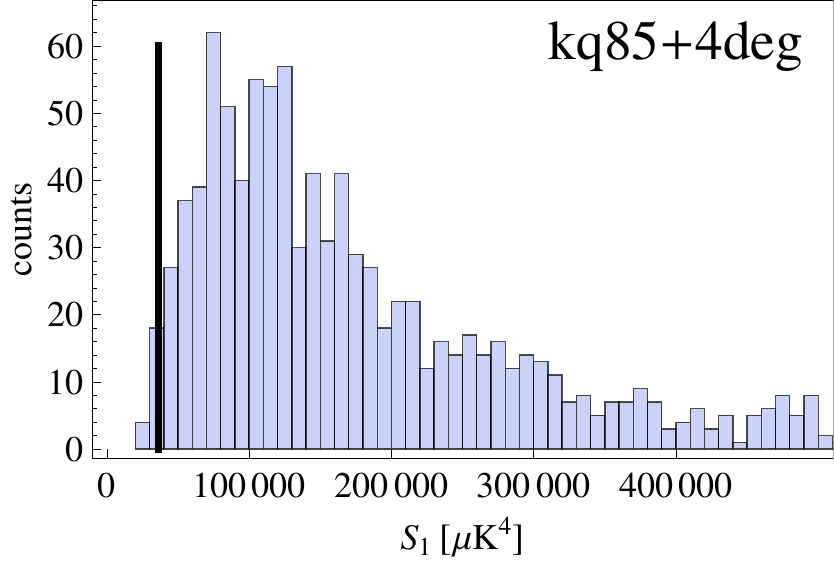}
\includegraphics[width=43mm]{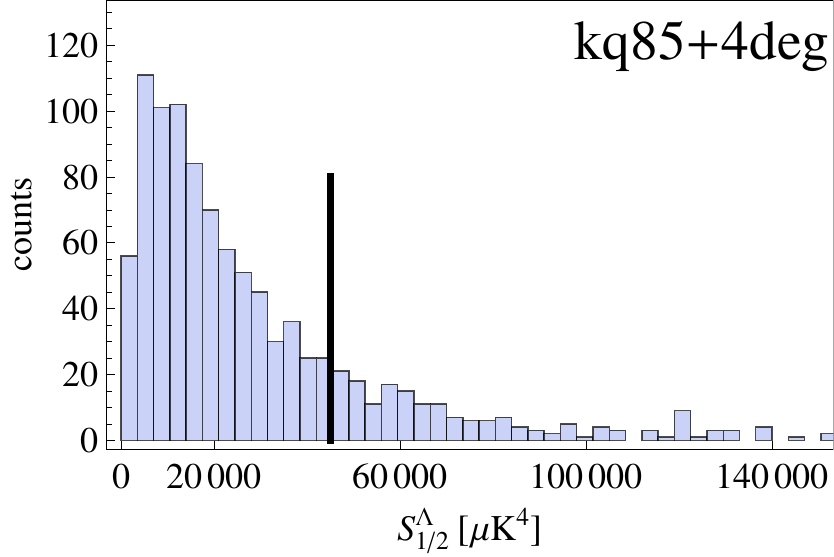}
\includegraphics[width=43mm]{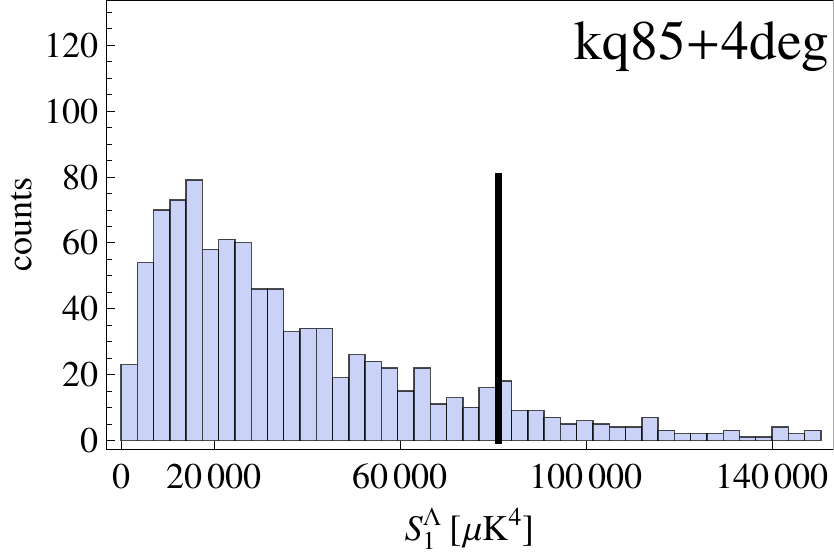}

\includegraphics[width=43mm]{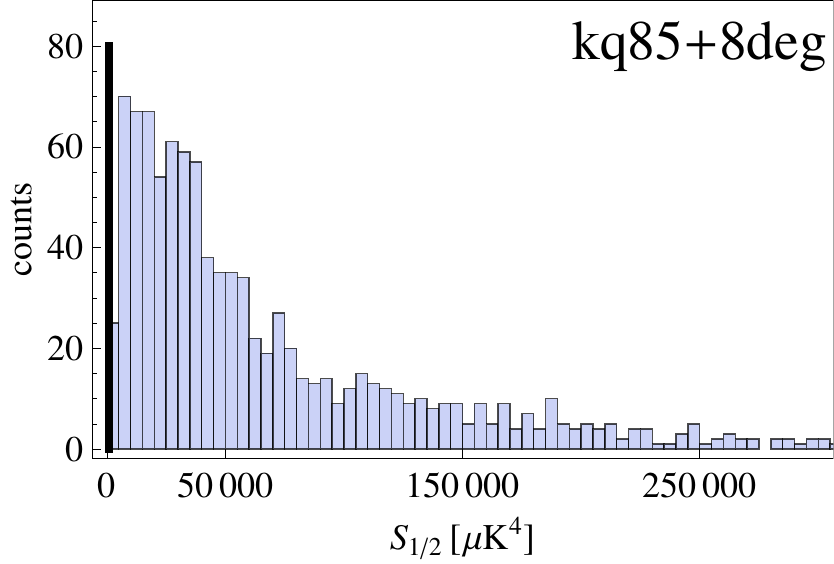}
\includegraphics[width=43mm]{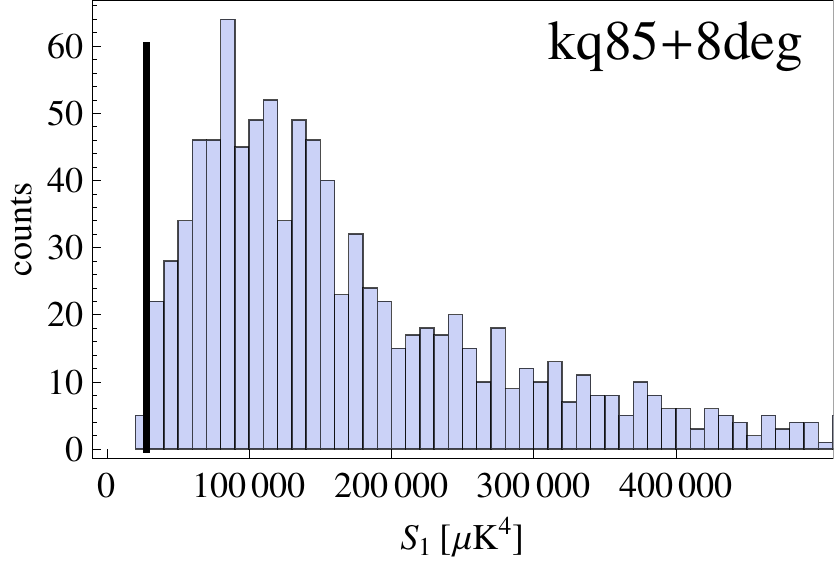}
\includegraphics[width=43mm]{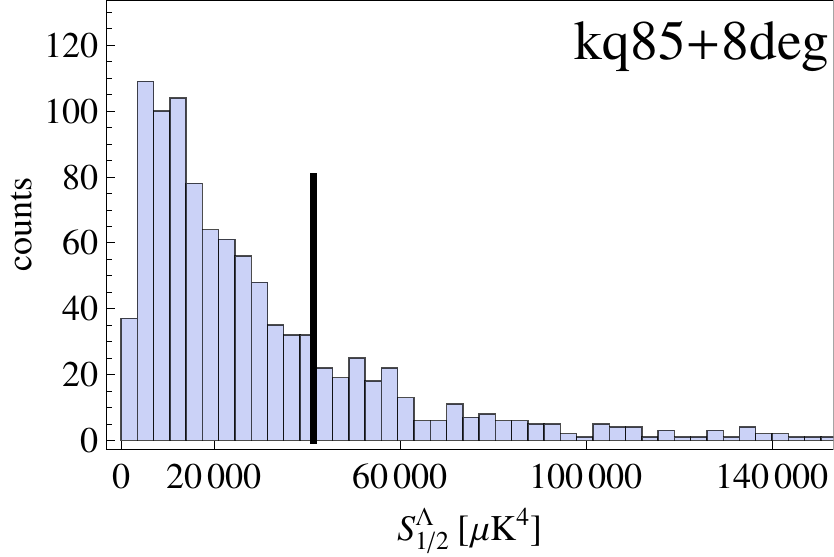}
\includegraphics[width=43mm]{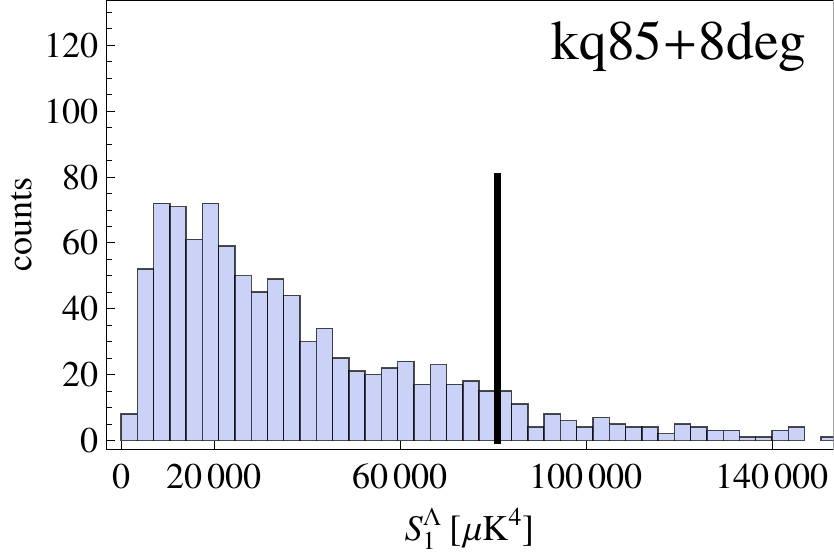}

\includegraphics[width=43mm]{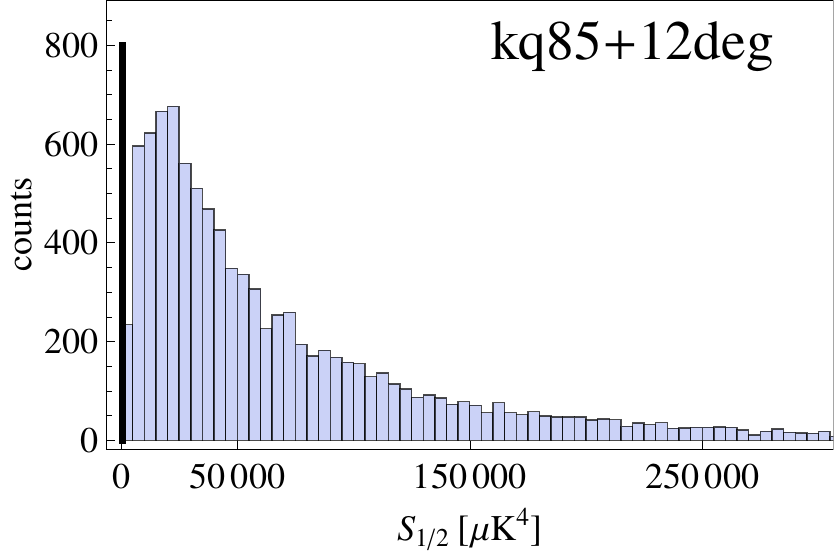}
\includegraphics[width=43mm]{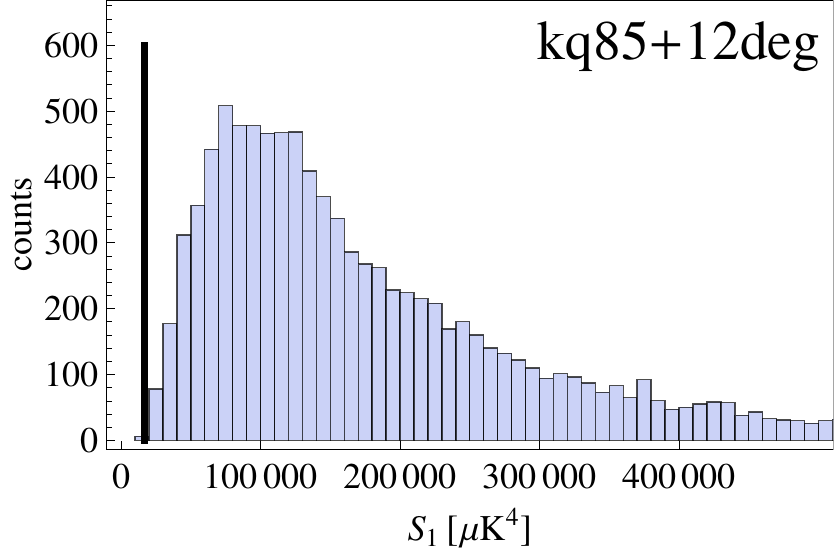}
\includegraphics[width=43mm]{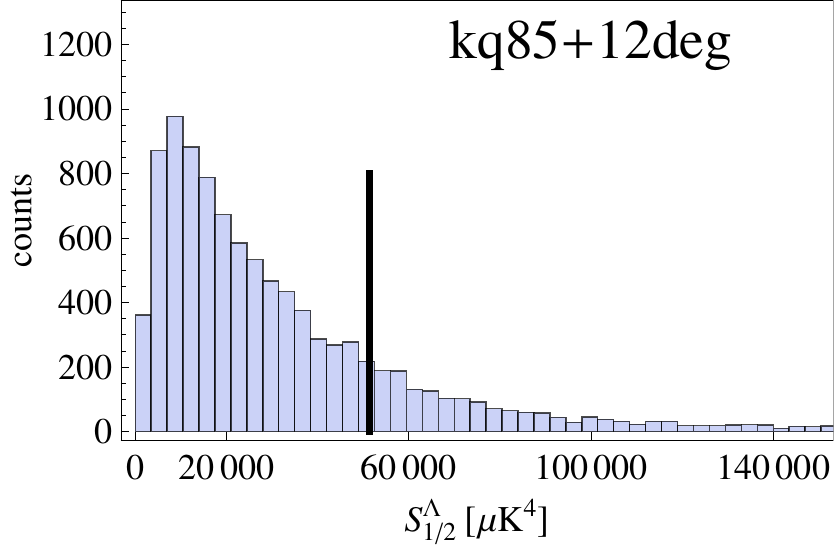}
\includegraphics[width=43mm]{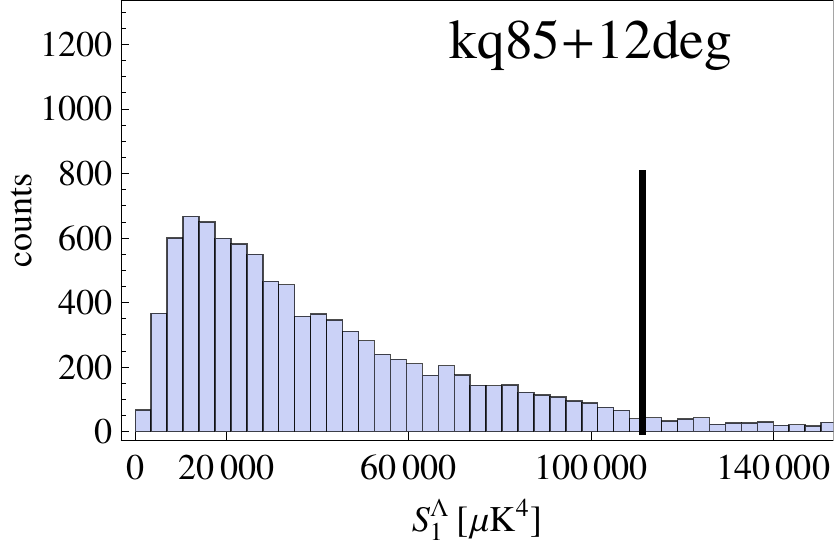}

\includegraphics[width=43mm]{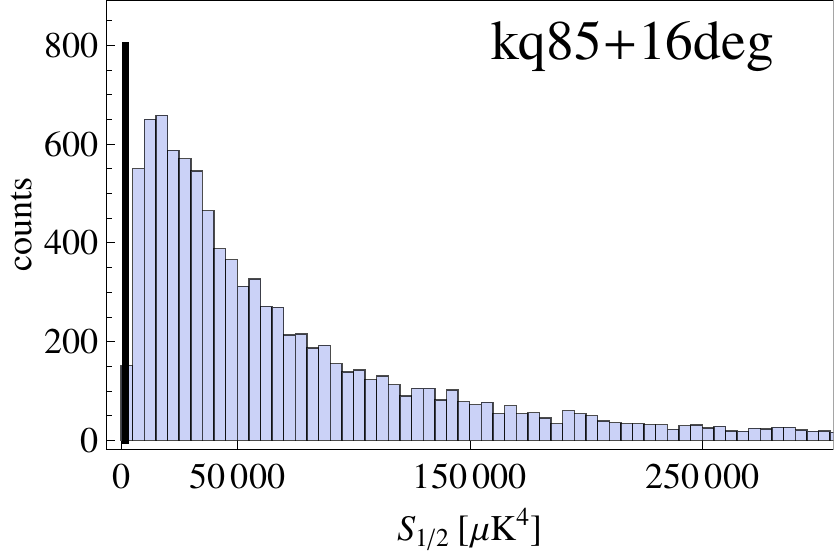}
\includegraphics[width=43mm]{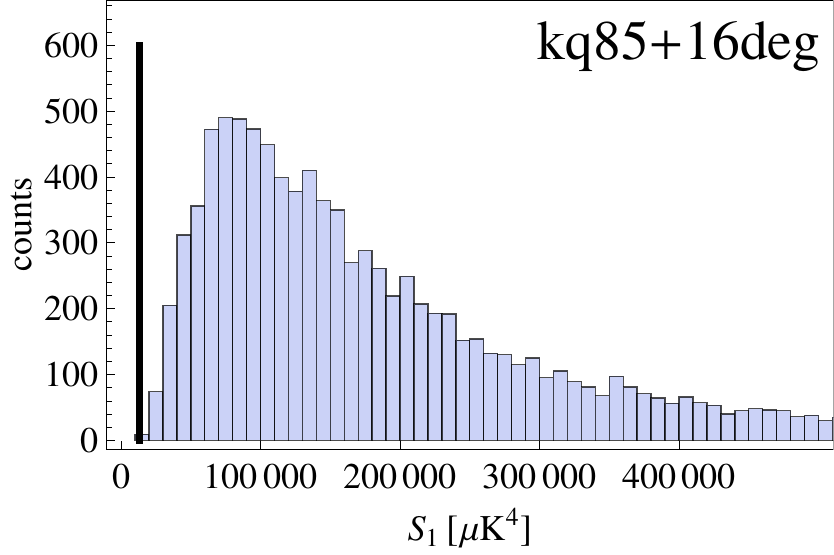}
\includegraphics[width=43mm]{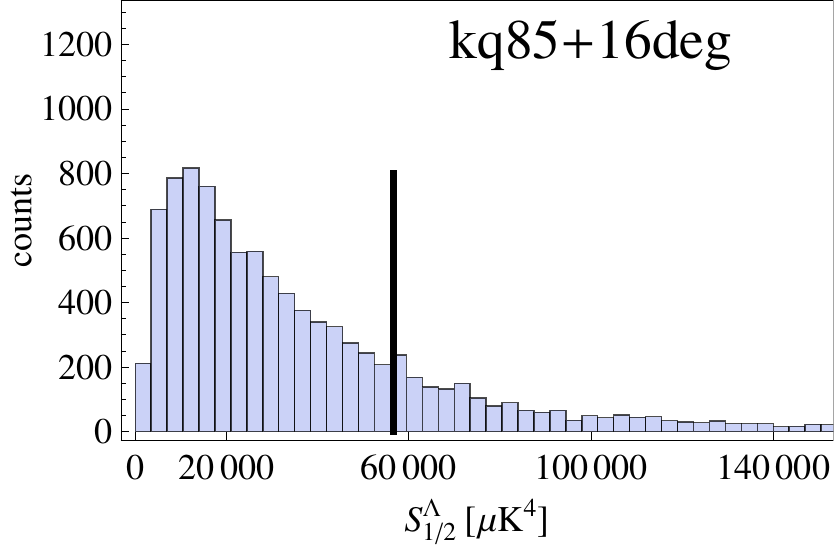}
\includegraphics[width=43mm]{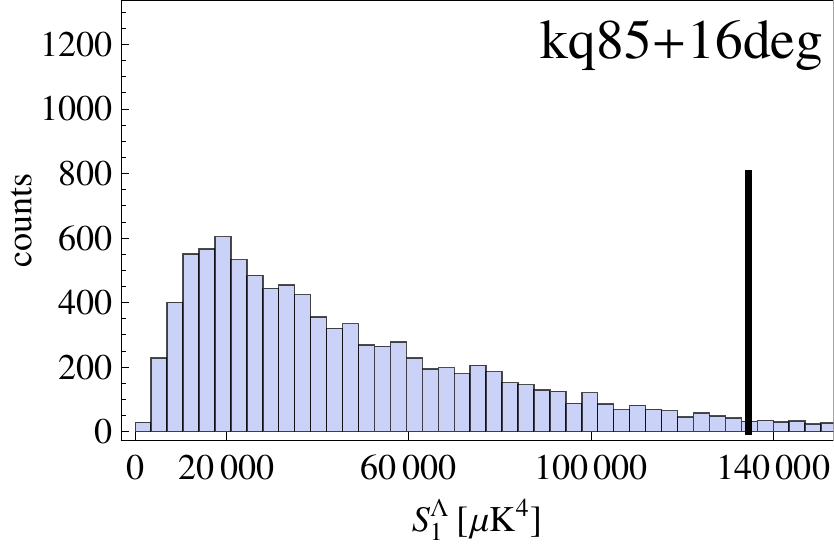}

\includegraphics[width=43mm]{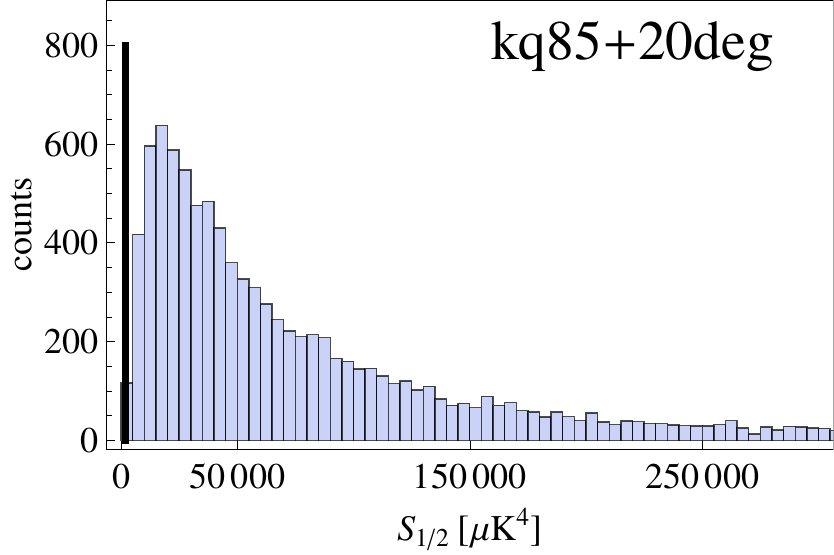}
\includegraphics[width=43mm]{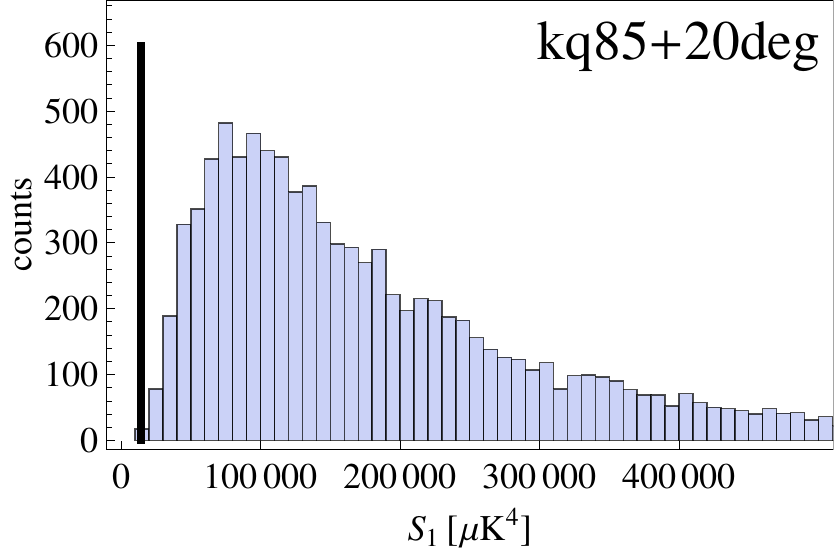}
\includegraphics[width=43mm]{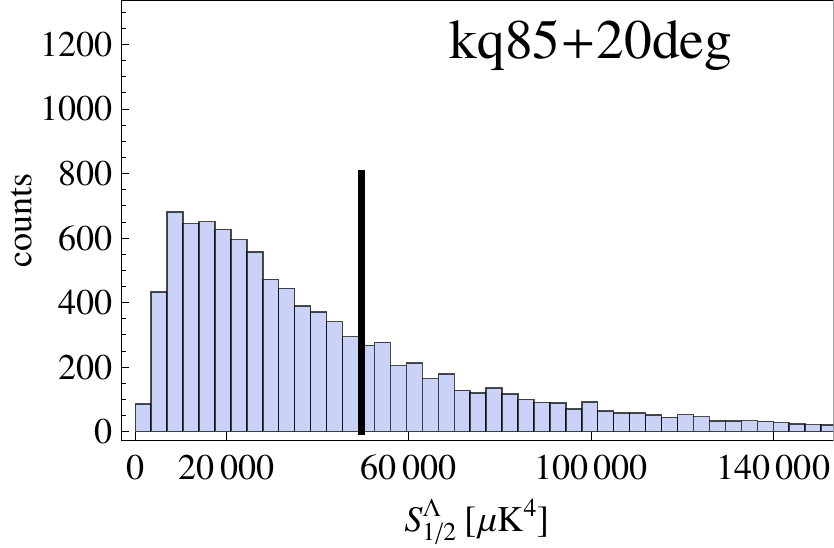}
\includegraphics[width=43mm]{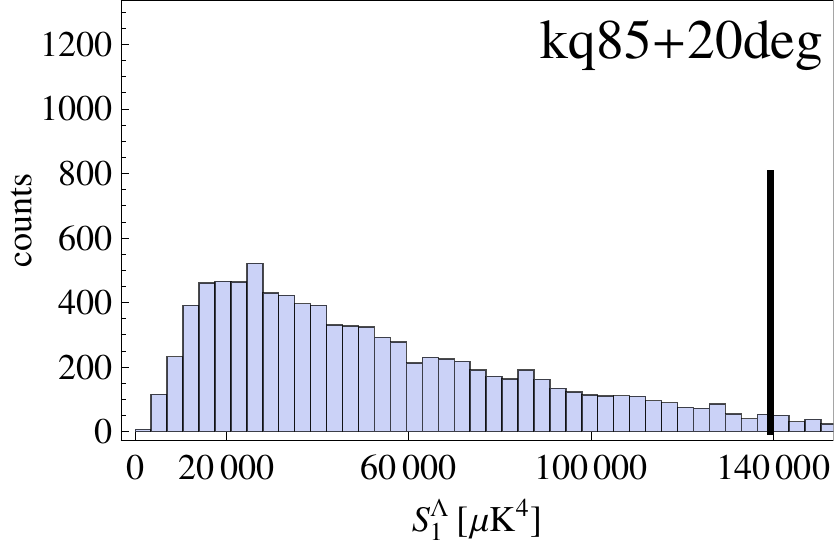}

\caption{Each panel represents the histogram of the estimators defined in Eqs.~(\ref{s1su2estimator}), (\ref{sfullestimator}), (\ref{s1su2estimatorLambda}) and (\ref{sfullestimatorLambda}). Units: counts (y-axis) versus the estimator $\mu$K$^4$ (x-axis) in all the panels. 
From left to right $S_{1/2}$, $S_{1}$, $S_{1/2}^{\Lambda}$ and $S_{1}^{\Lambda}$ are given. From upper to lower panels  ``a'', ``b'', ``c'', ``d'', ``e'' and ``f''  cases are shown.
These estimators are obtained replacing the TT APS into Eq.~(\ref{CTT}) with $\ell_{max}=32$. The thick solid line is for WMAP 9 year data.}
\label{quattro}
\end{figure*}

This Section represents the quantitative analysis of the paper.
Supported by realistic (signal plus noise) MC simulations we evaluate the estimators $S_{1/2}$, $S_{1}$, $S_{1/2}^{\Lambda}$, $S_{1}^{\Lambda}$ defined by Eqs.~(\ref{s1su2estimator}), (\ref{sfullestimator}),  (\ref{s1su2estimatorLambda}), (\ref{sfullestimatorLambda}) and compare with WMAP 9 year data. The histograms of the distributions of these estimators are given in Fig.~\ref{quattro} for each of the cases described in Table \ref{maskstabel}.
Each histogram (whose units are ``total counts'' vs $\mu$K$^4$) represents what is expected in $\Lambda$CDM model defined through the best fit of WMAP 9 data. 
The vertical bars are for the WMAP 9 year observations.

All the percentages of probability to obtain a value smaller than what observed by WMAP 9 are reported in Table \ref{percentages}.

\begin{figure}
\centering
\includegraphics[width=82mm]{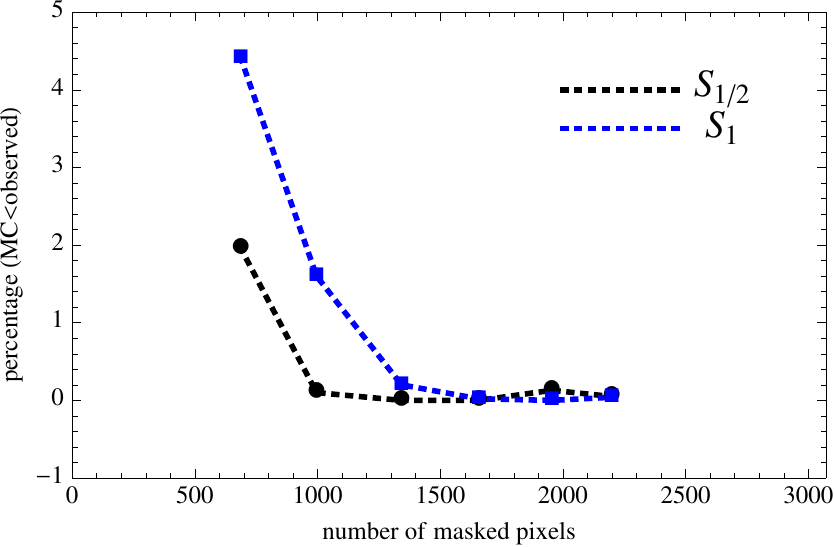}
\caption{
Percentage anomaly (i.e. lower tail probability) of $S_{1/2}$ and $S_{1}$ versus the number of masked pixels}
\label{quattrotris}
\end{figure}

We find that the behavior of the WMAP 9 observations are in general more compatible with $0$, i.e. no correlation, when the mask is enlarged.
This is quantified by the analysis of $S_{1/2}$ and $S_{1}$, see first and second column respectively of Figure \ref{quattro}.
For $S_{1/2}$, the probability to find an observed sky as the one provided by WMAP 9 can be very low, with a percentage less than $0.01 \%$
in the ``d'' case. 
Even for its generalization $S_{1}$, which is not suffering of any ``a posteriori bias'' since there is no arbitrary choice of the angular range over which perform the integration, 
we find an anomalous probability i.e. less than $0.01 \%$
for the case ``f''.
See Figure \ref{quattrotris} for a plot that shows the percentages of the anomaly of $S_{1/2}$ and $S_{1}$ versus the number of masked pixels.

We also find that the behavior of the WMAP 9 observations are in agreement with $\Lambda$CDM model, see third and fourth column of Figure \ref{quattro}
where the estimators $S_{1/2}^{\Lambda}$, $S_{1}^{\Lambda}$ are shown.

\begin{table}
\centering
\caption{Percentages to obtain a value smaller than what observed by WMAP 9. See also Fig.~\ref{one}.}
\label{percentages}
\begin{tabular}{cccccc}
\hline
case & $S_{1/2}$ & $S_{1}$ & $S_{1/2}^{\Lambda}$ & $S_{1}^{\Lambda}$ & $N_{sims}$\\
\hline
a & 1.96 & 4.41 & 80.77 & 84.84 & $10^4$\\
b & 0.1 &  1.6 & 78.9 & 85.4 & $10^3$ \\
c & $< 0.1$ & 0.2 & 74.7 & 85.1 & $10^3$ \\
d & $< 0.01$ & 0.02 & 79.35 & 90.13 & $10^4$ \\
e & $ 0.13$ & $<0.01$ & 77.48 & 90.56 & $10^4$ \\
f & $ 0.05$ & 0.04 & 66.29 & 88.64 & $10^4$ \\
\hline
\end{tabular}
\end{table}

See Appendix \ref{consistency} for a comparison between the estimators $S_{1/2}$, $S_{1}$, $S_{1/2}^{\Lambda}$ and $S_{1}^{\Lambda}$ build with the APS estimated by {\sc BolPol} and with the spectrum provided by the WMAP team  \citep{Bennett:2012zja}. This is done for case ``a''.

\section{Connection with APS}
\label{connectionwithAPS}

Since the shape of the two-point correlation function for $\theta > 60^{\circ}$ is dominated by the lowest harmonic modes, see Eq.~(\ref{CTT}), 
it is natural to link such lack of cross-correlation with the low amplitude of the lowest $C_{\ell}$, see for instance \citep{Gruppuso:2013xba}
where the APS is provided for the same data set and masks considered in Table \ref{maskstabel}.
In order to provide a quantitative analysis of this connection, we consider as an example, case ``d'', which is one of the most anomalous cases, see Table \ref{percentages},
and other two artificial cases, named ``d+C2'' and ``d+C2+C3''. 
The latter are defined starting from case ``d'' with the quadrupole replaced with the quadrupole value of the WMAP 9 best fit model (``d+C2'') and 
with the quadrupole and the octupole replaced by the values of the WMAP 9 best fit model (``d+C2+C3'').

\begin{figure}
\centering 
\includegraphics[width=80mm]{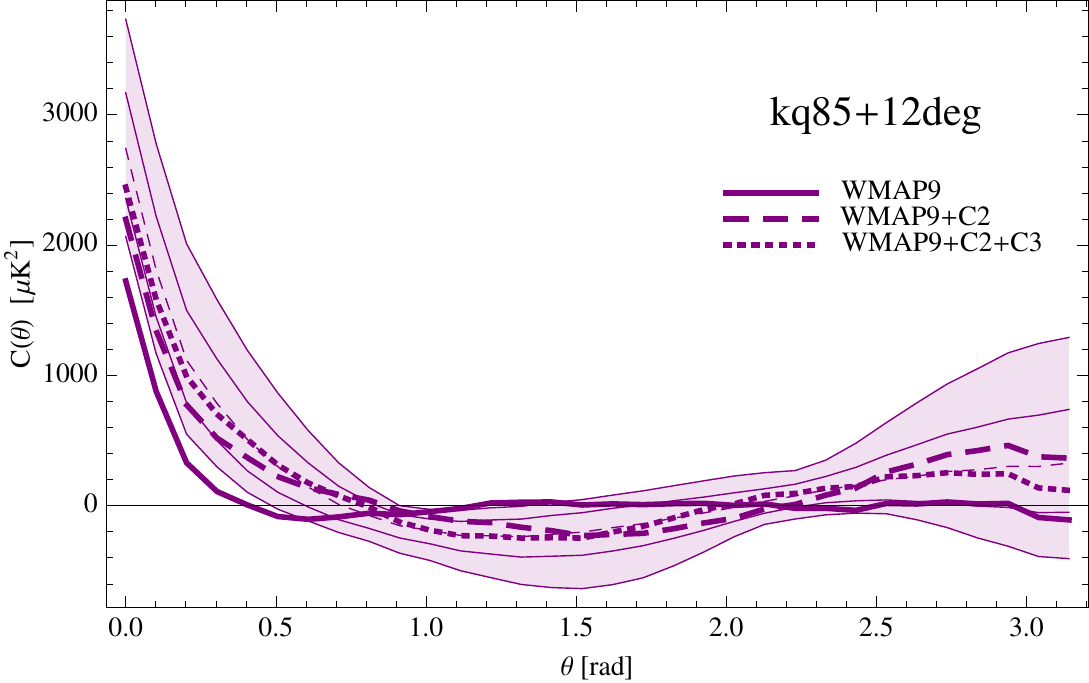}
\caption{TT two point correlation function for case ``d'' (solid line) and the artificial cases ``d+C2'' (dashed line) and ``d+C2+C3'' (dotted line).
Units: $\mu$K$^2$ (y-axis) and radiants (x-axis).}
\label{2point-fakeC2eC3} 
\end{figure}

\begin{figure}
\centering
\includegraphics[width=80mm]{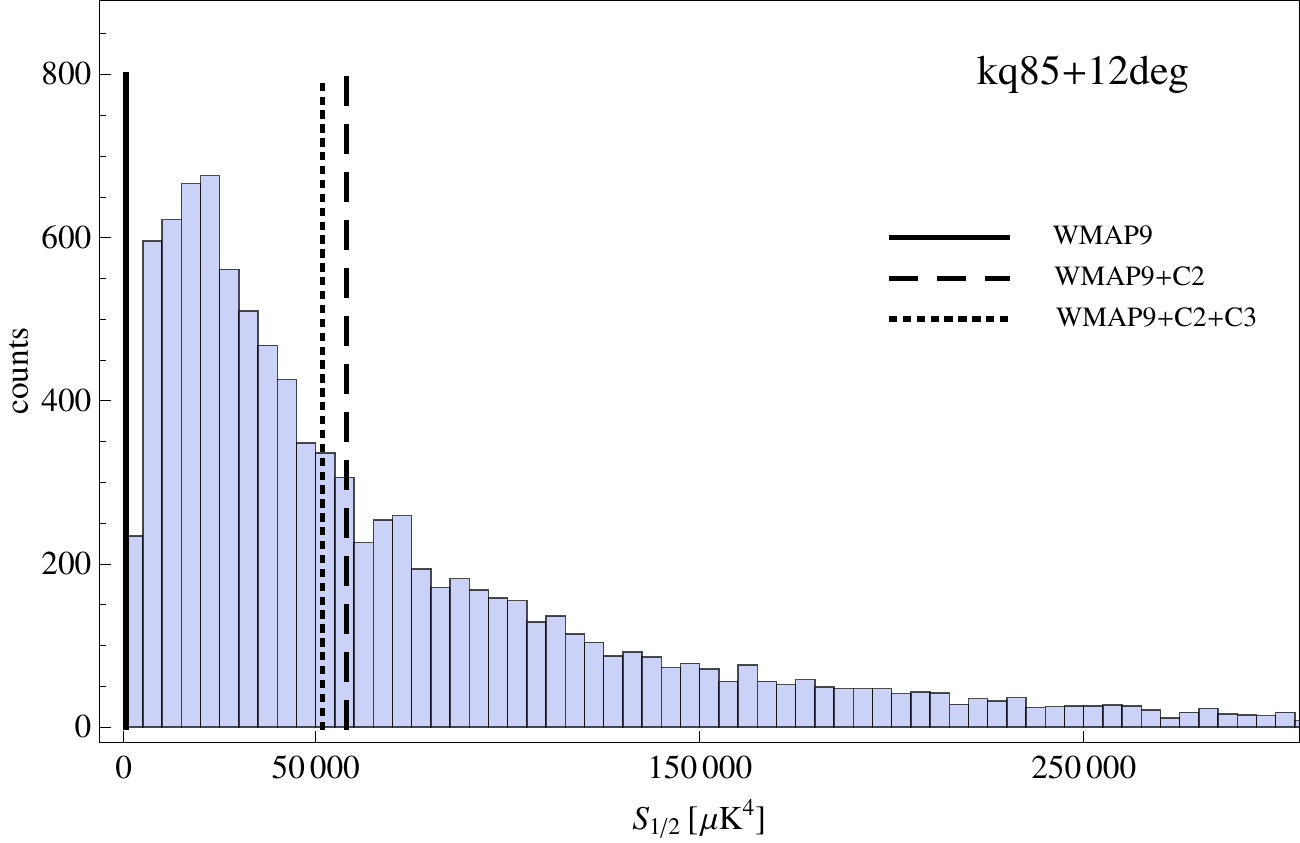}
\includegraphics[width=80mm]{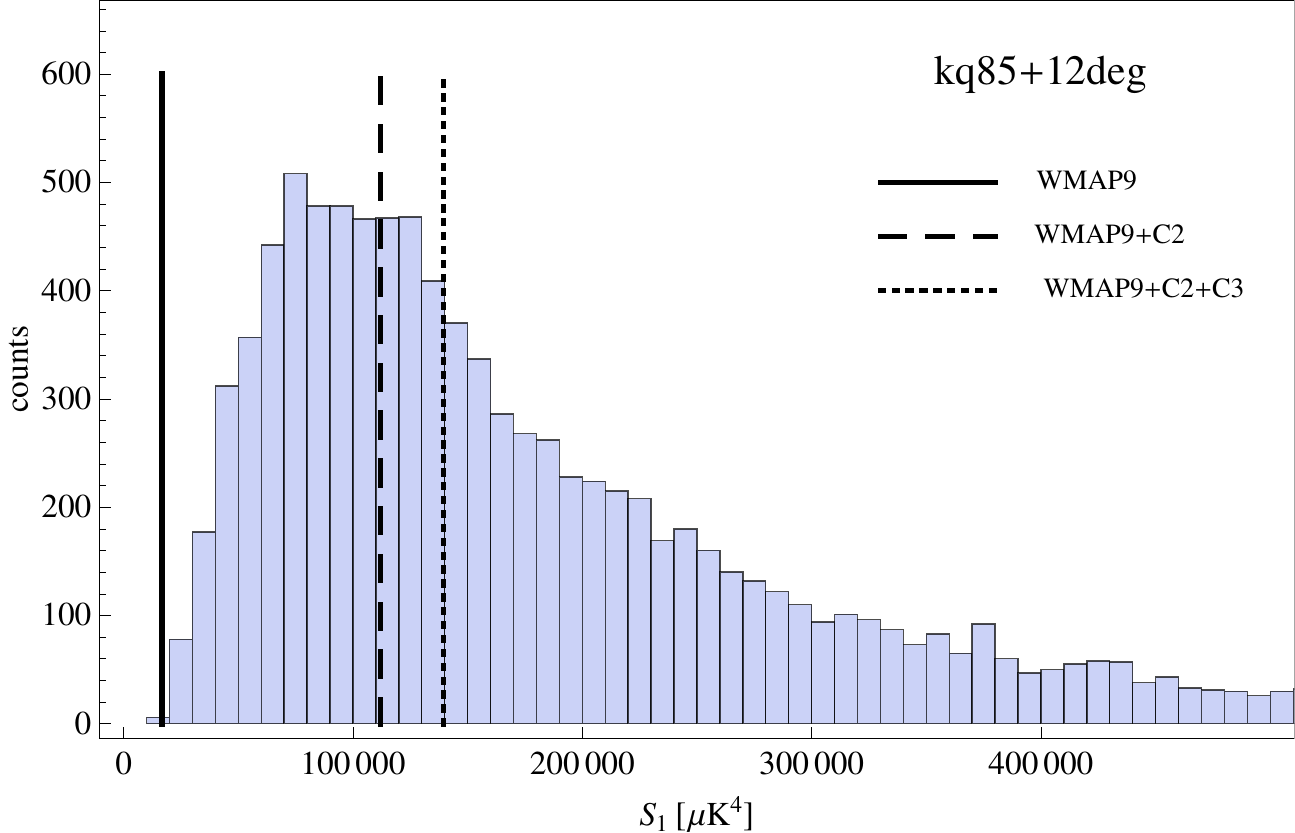}
\caption{In each panel the histogram represents the expected distribution of the estimator $S_{1/2}$ (see upper panel) and $S_{1}$ (see lower panel) in a $\Lambda$CDM model. 
Units: counts (y-axis) versus the estimator $\mu$K$^4$ (x-axis) in all the panels. 
The vertical bars stand for WMAP 9 value of the estimators for case ``d'' (solid line) and the artificial cases ``d+C2'' (dashed line) and ``d+C2+C3'' (dotted line).
These estimators are obtained replacing the TT APS into Eq.~(\ref{CTT}) with $\ell_{max}=32$.}
\label{S1su2eS1-fakeC2eC3} 
\end{figure}

In Fig.~\ref{2point-fakeC2eC3} we show the two-point correlation function for these new artificial cases ``d+C2'' (dashed line) and ``d+C2+C3'' (dotted line). 
For comparison we plot again the original two-point correlation function for case ``d'' (solid line).
Fig.~\ref{2point-fakeC2eC3} shows clearly that the two artificial cases are no more so close to the zero value as the original one.
To evaluate their distances from zero, we recompute the estimators $S_{1/2}$ and $S_1$, see Fig.~\ref{S1su2eS1-fakeC2eC3}, 
with the corresponding lower tail probabilities, see Table \ref{percentagesfake}. 
Since the new lower tail probabilities are much larger (they are around $50 \%$ compared to levels of $0.01 \%$, see Table \ref{percentagesfake}) 
it is not possible to talk about anomaly anymore. 
This shows that the lack of correlation is driven by the low amplitude of the lowest multipoles.
This also indicate that the so called ``Low Variance'' at large angular scales, studied in  \citep{Gruppuso:2013xba} with the same data set, has the same origin as the lack 
of correlation in the two-point correlation function.

\begin{table}
\centering
\caption{Percentages to obtain a value smaller than what observed by WMAP 9 for case ``d'' and the artificial ``d+C2'' and ``d+C2+C3'' cases. See also the text.}
\label{percentagesfake}
\begin{tabular}{cccc}
\hline
case & $S_{1/2}$ & $S_{1}$ & $N_{sims}$\\
\hline
d & $< 0.01$ & 0.02 & $10^4$ \\
d+C2 & $ 56.29$ & $ 34.08 $  & $10^4$ \\
d+C2+C3 & $ 52.21$ & 46.28 &  $10^4$ \\
\hline
\end{tabular}
\end{table}

\section{Conclusions}
\label{conclusions}

In the present paper we have evaluated the two-point correlation function of WMAP 9 year data.  
This function has been computed using the APS estimates obtained through a QML method which is proven to be optimal \citep{Gruppuso:2009ab},
see also Appendix \ref{QMLBolpol}.
The behavior of this function has been tested against various Galactic masks, see Table \ref{maskstabel} and Fig.~\ref{one}.
The cases of Table \ref{maskstabel} have been confronted with the WMAP 9 best fit model, see Fig.~\ref{fig:minipage2}.
This has been possible thanks to MC realistic simulations, each of them being analyzed with our implementation of QML estimator.
Looking at Fig.~\ref{fig:minipage2}, we have qualitatively noted that the increase of the mask pushes $C(\theta)$ downward for $\theta < 60^{\circ}$. 
At the same time at large scales, i.e. $\theta > 150^{\circ}$, $C(\theta)$ is going systematically upward still when the mask is larger. This means that $C(\theta)$
is more consistent with $\Lambda$CDM model (and with no-correlation) when the sky area around the kq85 mask is dropped out from the analysis. 
The latter behavior also suggests that increasing the mask, the so called TT Parity anomaly is becoming milder, 
since $C(\theta=\pi)$ is a natural estimator for the even-odd multipole power asymmetry.

Moreover we have quantitatively evaluated  $S_{1/2}$ and  $S_{1}$ in order to estimate the consistency with no correlation.
We have demonstrated that the anomaly is showing up enlarging the kq 85 mask.
Fig.~\ref{quattrotris} makes evident how the anomalies percentage monotonically decreases with the masked area.
Note that not only $S_{1/2}$ is anomalous for a $\Lambda$CDM model (see first column of Table \ref{percentages}) but also its generalization $S_1$,
which does not suffer of any ``a posteriori bias'', is very unlikely at the level of $ \lesssim 0.04\%$ C.L. for the considered masks that 
cover at least $\sim 54\%$ of the sky (see second column of Table \ref{percentages}).
The other two considered estimators, $S_{1/2}^{\Lambda}$ and  $S_{1}^{\Lambda}$ are found to be consistent with $\Lambda$CDM model, 
see third and fourth columns of Table \ref{percentages}.
See Fig.~\ref{quattro} for an explicit computation of the distribution of $S_{1/2}$, $S_{1}$, $S_{1/2}^{\Lambda}$ and  $S_{1}^{\Lambda}$ 
in a $\Lambda$CDM model (histograms) and the corresponding WMAP value (vertical bars).

Furthermore we have shown that increasing artificially the quadrupole and octupole values from what observed \citep{Gruppuso:2013xba} to the WMAP 9 best fit values, 
makes $S_{1/2}$ and  $S_{1}$ not anomalous, see Fig.~\ref{S1su2eS1-fakeC2eC3} and Table \ref{percentagesfake}. 
This indicates that the low amplitude of the lowest APS are responsible of the lack of correlation in the two-point correlation function.
This fact represents a connection with the Low Variance anomaly which is driven by the same multipoles  \citep{Monteserin:2007fv,Cruz:2010ud,Gruppuso:2013xba}.

We will return to these analyses using { Planck} data in the near future.

\section*{Acknowledgments}
A.G. wishes to thank all the other developers of the {\sc BolPol} code, whose use is acknowledged here \citep{Gruppuso:2009ab}.
We acknowledge the use of computing facilities at NERSC (USA) and CINECA (ITALY). 
We acknowledge use of the HEALPix (Gorski et al. 2005) software and analysis package for deriving the results in this paper. 
We acknowledge the use of the Legacy Archive for Microwave Background Data Analysis (LAMBDA). 
Support for LAMBDA is provided by the NASA Office of Space Science. 
Work supported by ASI through ASI/INAF Agreement I/072/09/0 for the Planck LFI Activity of Phase E2 and by MIUR through PRIN 2009 (grant n. 2009XZ54H2).

\appendix

\section{QML estimator}
\label{QMLBolpol}
In order to evaluate the APS we adopt the QML estimator,
introduced in \citep{Tegmark:1996qt} and extended to polarization in \citep{Tegmark:2001zv}. 
In this appendix we describe the essence of such a method. 
Further details about the considered implementation can be found in \citep{Gruppuso:2009ab}.

Given a CMB temperature map, ${\bf x}$, the QML provides estimates
$\hat {C}_\ell$ of the APS as: 
\begin{equation}
{\ell (\ell +1) \over 2 \pi} \hat{C}_\ell = \sum_{\ell'} (F^{-1})_{\ell\ell'} \left[ {\bf x}^t
{\bf E}^{\ell'} {\bf x}-tr({\bf N}{\bf
E}^{\ell'}) \right]
\, ,
\end{equation}
where the $F^{\ell \ell '}$ is the Fisher matrix, defined as
\begin{equation}
\label{eq:fisher}
F^{\ell\ell'}= \mu_{\ell}  \mu_{\ell'} \frac{1}{2}tr\Big[{\bf C}^{-1}\frac{\partial
{\bf C}}{\partial
  C_\ell}{\bf C}^{-1}\frac{\partial {\bf C}}{\partial
C_{\ell'}}\Big] \,,
\end{equation}
and the ${\bf E}^{\ell}$ matrix is given by
\begin{equation}
\label{eq:Elle}
{\bf E}^\ell=  \mu_{\ell}  \frac{1}{2}{\bf C}^{-1}\frac{\partial {\bf C}}{\partial
  C_\ell}{\bf C}^{-1} \, ,
\end{equation}
with ${\bf C} ={\bf S}(C_{\ell})+{\bf N}$ being the global covariance matrix (signal plus noise contribution)
and  $\mu_{\ell} = { 2 \pi b_{\ell}^2 / \ell (\ell +1)}$, where the $b_{\ell}$ are the beam window function including the pixel window function. 

Although an initial assumption for a fiducial power spectrum $C_{\ell}$ is needed, the QML method provides unbiased estimates of the power spectrum contained 
in the map regardless of the initial guess,
\begin{equation}
\langle\hat{C}_\ell \rangle= \bar C_\ell \,,
\label{unbiased}
\end{equation}
where the average is taken over the ensemble of realizations (or, in a practical test, over Monte Carlo 
realizations extracted from $\bar C_\ell $).
On the other hand, the covariance matrix associated to the estimates,
\begin{equation}
\langle\Delta\hat{C}_\ell
\Delta\hat{C}_{\ell'} \rangle= \mu_{\ell} \mu_{\ell'}( F^{-1})_{\ell\ell'} \,,
\label{minimum}
\end{equation}
does depend on the initial assumption for $C_\ell$: 
the closer the guess to the true power spectrum is, the closer are the error bars to minimum variance.
According to the Cramer-Rao inequality, which sets a limit to the accuracy of an estimator, Eq. (\ref{minimum}) tells us that 
the QML has the smallest error bars.  The QML is then an `optimal' estimator because it saturates the Cramer-Rao bound.

We have tested that this is the case for our QML implementation, i.e. {\sc BolPol}. This has been checked under the assumption 
of Gaussianity of CMB anisotropies.

\section{Consistency with WMAP 9 results}
\label{consistency}

We provide here a comparison of the two-point correlation function and of estimators $S_{1/2}$, $S_{1}$, $S_{1/2}^{\Lambda}$, $S_{1}^{\Lambda}$ 
computed with the {\sc BolPol} spectrum and with the publicly available spectrum provided by the WMAP team \citep{Bennett:2012zja}.
The latter is obtained maximizing the likelihood distribution at a given multipole, fixing the others to the WMAP 9 best fit model, up to and including $\ell_{max}=32$. 

\begin{figure}
\centering 
\includegraphics[width=80mm]{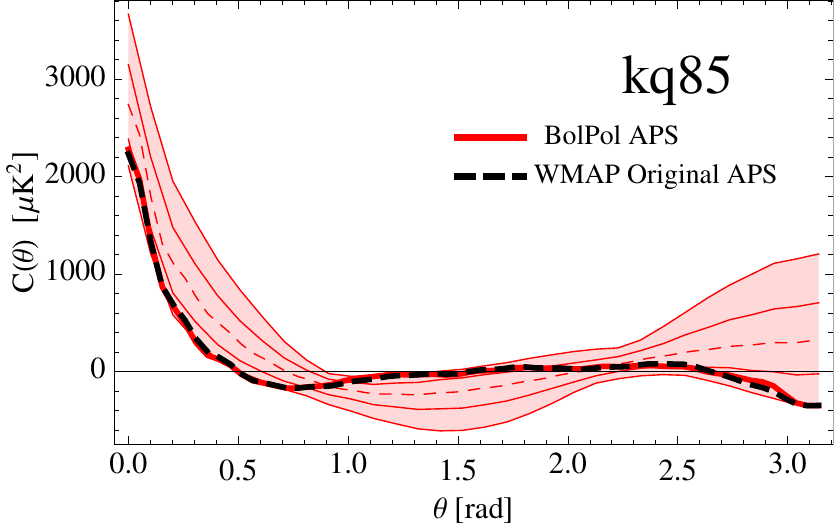}
\caption{TT two point correlation function for case ``a''. Red solid line is for the function computed with the {\sc BolPol} spectrum and black dashed line for the function 
computed with the original WMAP spectrum.
Units: $\mu$K$^2$ (y-axis) and radiants (x-axis).}
\label{2point-comparison} 
\end{figure}

\begin{figure}
\centering
\includegraphics[width=41mm]{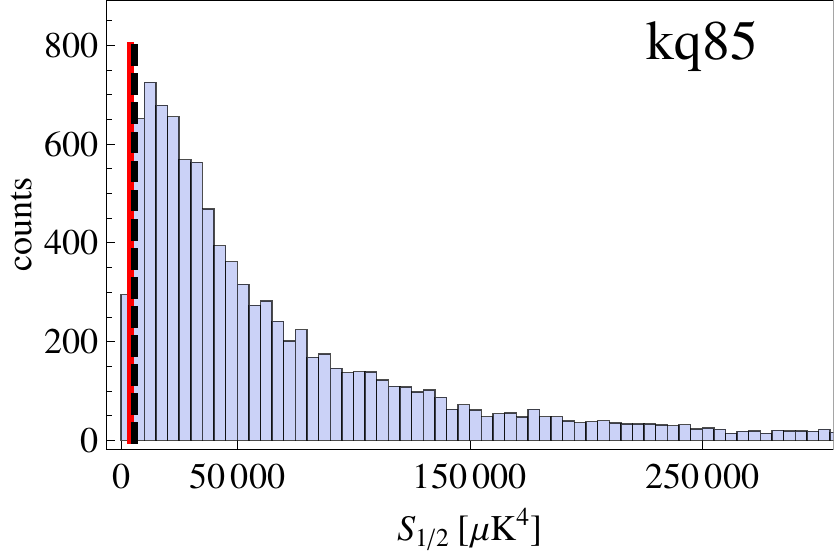}
\includegraphics[width=41mm]{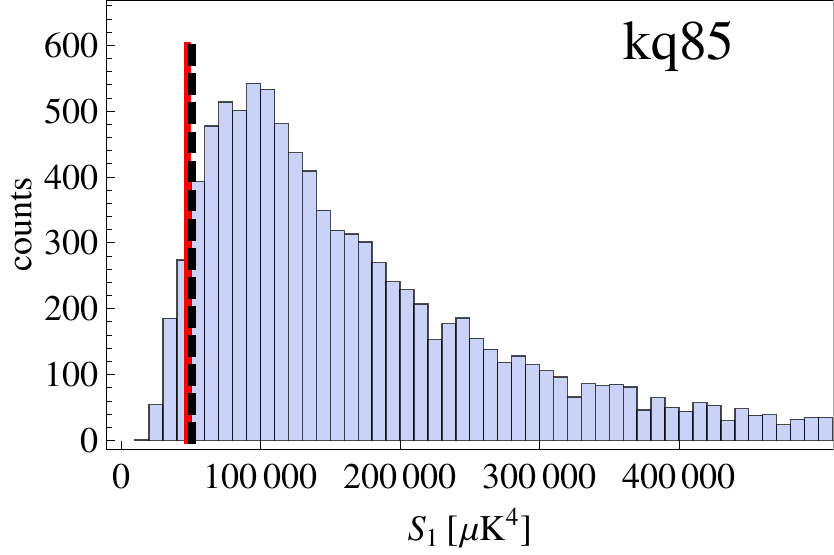}
\includegraphics[width=41mm]{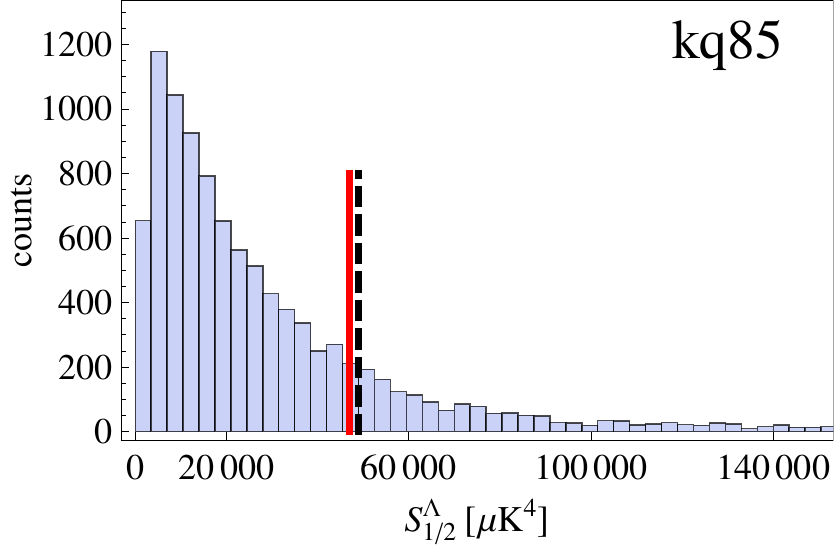}
\includegraphics[width=41mm]{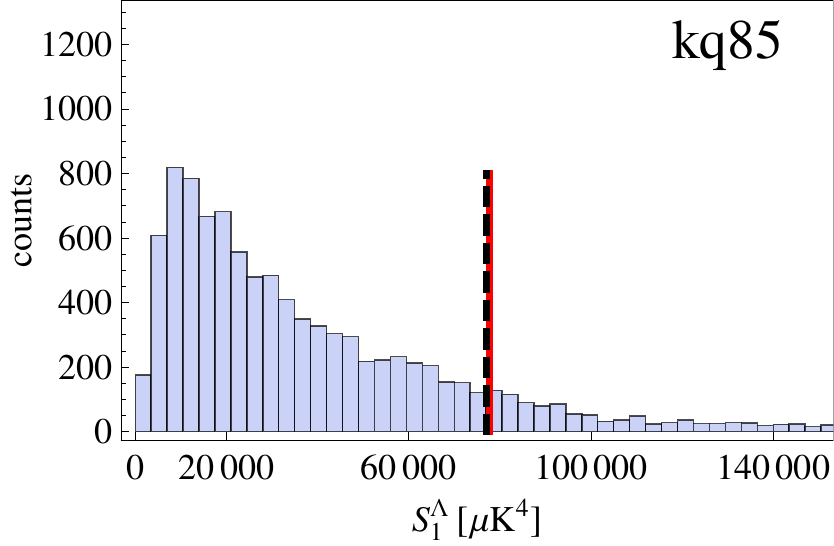}
\caption{In each panel the histogram represents the expected distribution of the estimator $S_{1/2}$ (upper left panel), $S_{1}$ (upper right panel),
$S_{1/2}^{\Lambda}$ (lower left panel), $S_{1}^{\Lambda}$ (lower right panel)
in a $\Lambda$CDM model. 
Units: counts (y-axis) versus the estimator $\mu$K$^4$ (x-axis) in all the panels. 
Red solid line is for the estimators computed with the {\sc BolPol} spectrum and black dashed line for the estimators
computed with the original WMAP spectrum.
All the  estimators are obtained replacing the TT APS into Eq.~(\ref{CTT}) with $\ell_{max}=32$.}
\label{Estimators-comparison} 
\end{figure}

In Fig.~\ref{2point-comparison} we show the two-point correlation function for case ``a''.
Red solid line and black dashed lines are obtained replacing in Eq.~(\ref{CTT}) the spectrum obtained by {\sc BolPol} and by the WMAP team respectively, up to $\ell_{max}=32$.
In Fig.~\ref{Estimators-comparison} we show the estimators $S_{1/2}$, $S_{1}$, $S_{1/2}^{\Lambda}$, $S_{1}^{\Lambda}$ for case ``a''.
Red solid and black dashed vertical bars are obtained with the {\sc BolPol} spectrum and the original WMAP spectrum respectively.

Fig.~\ref{2point-comparison} and Fig.~\ref{Estimators-comparison} show that there is a good consistency between the two spectra, the one obtained 
through {\sc BolPol} and the one provided by the WMAP team.



\begin{thebibliography}{99}

\bibitem[\protect\citeauthoryear{Bennett et al.}{2010}]{Bennett:2010jb} 
  Bennett C.~L. et al., 2011,
  Astrophys.\ J.\ Suppl.\  {192}, 17 
  
\bibitem[\protect\citeauthoryear{Bennett et al.}{2012}]{Bennett:2012zja}
  Bennett C.~L. {et al.}, 2012,  
  Accepted for publication to Astrophys.\ J.\ Suppl. 
  arXiv:1212.5225 [astro-ph.CO].

  
\bibitem[\protect\citeauthoryear{Copi et al.}{2007}]{Copi:2006tu} 
  Copi C., Huterer D., Schwarz D., Starkman G., 2007,
  Phys.\ Rev.\ D {75}, 023507 
  
\bibitem[\protect\citeauthoryear{Copi et al.}{2009}]{Copi:2008hw} 
  Copi C.~J., Huterer D., Schwarz D.~J., Starkman G.~D., 2009,
  Mon.\ Not.\ Roy.\ Astron.\ Soc.\  {399}, 295 
  
\bibitem[\protect\citeauthoryear{Copi et al.}{2010}]{Copi:2010na} 
  Copi C.~J., Huterer D., Schwarz D.~J., Starkman G.~D., 2010,
  Adv.\ Astron.\  {2010}, 847541 
  
\bibitem[\protect\citeauthoryear{Copi et al.}{2013}]{Copi:2013zja} 
  Copi C.~J., Huterer D., Schwarz D.~J., Starkman G.~D., 2013,
  preprint (arXiv:1303.4786) 

\bibitem[\protect\citeauthoryear{Cruz et al.}{2011}]{Cruz:2010ud} 
  Cruz M., Vielva P., Martinez-Gonzalez E., Barreiro R.~B., 2011,
Mon.\ Not.\ Roy.\ Astron.\ Soc.\  {412}, 2383
%
%
%
  
\bibitem[\protect\citeauthoryear{Dunkley et al.}{2009}]{Dunkley:2008ie}
  Dunkley J.~ {et al.} [WMAP Collaboration], 2009,
  Astrophys.\ J.\ Suppl.\  {180}, 306-329 
  
\bibitem[\protect\citeauthoryear{Efstathiou}{2004}]{Efstathiou:2003tv} 
  Efstathiou G., 2004,
  Mon.\ Not.\ Roy.\ Astron.\ Soc.\  {348}, 885 
  
  
\bibitem[\protect\citeauthoryear{Efstathiou, Ma \& Hanson}{2010}]{Efstathiou:2009di} 
 Efstathiou G., Ma Y.~-Z., Hanson D., 2010, 
  Mon.\ Not.\ Roy.\ Astron.\ Soc.\  {407}, 2530
  arXiv:0911.5399 [astro-ph.CO].

\bibitem[\protect\citeauthoryear{Gorski et al.}{2005}]{gorski}
Gorski K.M., Hivon E., Banday A.J., Wandelt B.D., Hansen F.K., Reinecke M., Bartelmann M., 2005,
Ap.J., 622, 759-771


\bibitem[\protect\citeauthoryear{Gruppuso et al.}{2009}]{Gruppuso:2009ab}
  Gruppuso A., De Rosa  A., Cabella P., Paci F., Finelli F., Natoli P., de Gasperis G., Mandolesi N., 2009,
  Mon.\ Not.\ Roy.\ Astron.\ Soc.\  {400}, 463-469 
%

\bibitem[\protect\citeauthoryear{Gruppuso et al.}{2011}]{Gruppuso:2010nd} 
  Gruppuso A., Finelli F., Natoli P., Paci F., Cabella P., De Rosa A., Mandolesi N., 2011,
  Mon.\ Not.\ Roy.\ Astron.\ Soc.\  {411}, 1445 
  
\bibitem[\protect\citeauthoryear{Gruppuso et al.}{2012}]{Gruppuso:2011ci} 
  Gruppuso A., Natoli P., Mandolesi N., De Rosa A., Finelli F., Paci F., 2012,
  JCAP {1202}, 023 
  
\bibitem[\protect\citeauthoryear{Gruppuso et al.}{2013}]{Gruppuso:2013xba} 
  Gruppuso A., Natoli P., Paci F., Finelli F., Molinari D., De Rosa A., Mandolesi N., 2013,
  JCAP, in press 
  
\bibitem[\protect\citeauthoryear{Hinshaw et al.}{1996}]{Hinshaw:1996ut} 
  Hinshaw G., Banday A.~J., Bennett C.~L., Gorski K.~M., Kogut A., Lineweaver C.~H., Smoot G.~F., Wright E.~L., 1996,
  Astrophys.\ J.\  {464}, L25 

\bibitem[\protect\citeauthoryear{Hinshaw et al.}{2013}]{Hinshaw:2012fq} 
  Hinshaw G. et al., 2013,
  ApJS, in press 
  
%
  
\bibitem[\protect\citeauthoryear{Kim \& Naselsky}{2010a}]{Kim:2010gf} 
  Kim J., Naselsky P., 2010a,
  Astrophys.\ J.\  {714}, L265 
  
\bibitem[\protect\citeauthoryear{Kim \& Naselsky}{2010b}]{Kim:2010gd} 
  Kim J., Naselsky P., 2010b,
  Phys.\ Rev.\ D {82}, 063002 
  
\bibitem[\protect\citeauthoryear{Kim \& Naselsky}{2011}]{Kim:2010st} 
  Kim J., Naselsky P., 2011,
  Astrophys.\ J.\  {739}, 79 

\bibitem[\protect\citeauthoryear{Larson et al.}{2010}]{Larson:2010gs}
  Larson D. et al., 2011,
  Astrophys.\ J.\ Suppl.\  {192}, 16
  
  \bibitem[\protect\citeauthoryear{Molinari et al.}{2013}]{Molinari2013} 
  Molinari D. et al., 2013, to appear
  
\bibitem[\protect\citeauthoryear{Monteserin et al.}{2008}]{Monteserin:2007fv} 
  Monteserin C., Barreiro R.~B.~B., Vielva P., Martinez-Gonzalez E., Hobson M.~P., Lasenby A.~N., 2008,
  Mon.\ Not.\ Roy.\ Astron.\ Soc.\  {387}, 209 
   
\bibitem[\protect\citeauthoryear{Paci et al.}{2010}]{Paci:2010wp} 
  Paci F., Gruppuso A., Finelli F., Cabella P., De Rosa A., Mandolesi N., Natoli P., 2010,
  Mon.\ Not.\ Roy.\ Astron.\ Soc.\  {407}, 399
  
\bibitem[\protect\citeauthoryear{Paci et al.}{2013}]{Paci:2013gs} 
  Paci F., Gruppuso A., Finelli F., De Rosa A., Mandolesi N. and Natoli P., 2013,
  MNRAS, in press 
  
\bibitem[\protect\citeauthoryear{Planck Collaboration I}{2013}]{Ade:2013xsa} 
  Planck Collaboration, 2013,
  preprint (arXiv:1303.5062) 
  
\bibitem[\protect\citeauthoryear{Planck Collaboration XV}{2013}]{Planck:2013kta} 
  Planck Collaboration, 2013,
  preprint (arXiv:1303.5075) 
  
\bibitem[\protect\citeauthoryear{Planck Collaboration XVI}{2013}]{Ade:2013lta} 
  Planck Collaboration, 2013,
  preprint (arXiv:1303.5076) 
  
\bibitem[\protect\citeauthoryear{Planck Collaboration XX}{2013}]{Ade:2013lmv} 
  Planck Collaboration, 2013,
  preprint (arXiv:1303.5080) 

\bibitem[\protect\citeauthoryear{Planck Collaboration XXIII}{2013}]{Ade:2013sta} 
  Planck Collaboration, 2013,
  preprint (arXiv:1303.5083) 
  
%
  
\bibitem[\protect\citeauthoryear{Sarkar et al.}{2011}]{Sarkar:2010yj} 
  Sarkar D., Huterer D., Copi C.~J., Starkman G.~D. and Schwarz D.~J., 2011,
  Astropart.\ Phys.\  {34}, 591 

  
\bibitem[\protect\citeauthoryear{Spergel et al.}{2003}]{Spergel:2003cb} 
  Spergel D.~N. {et al.}  [WMAP Collaboration], 2003,
  Astrophys.\ J.\ Suppl.\  {148}, 175
  
\bibitem[\protect\citeauthoryear{Tegmark}{1997}]{Tegmark:1996qt} 
  Tegmark M., 1997,
  Phys.\ Rev.\ D {55}, 5895 
  
\bibitem[\protect\citeauthoryear{Tegmark \& de Oliveria-Costa}{2001}]{Tegmark:2001zv}
  Tegmark M., de Oliveira-Costa A., 2001,
  Phys.\ Rev.\ D {64} 063001

\end{thebibliography}
\end{document}